\pdfoutput=1
\documentclass{jfm}
\usepackage{graphicx}
\usepackage{epstopdf, epsfig}
\usepackage{changes}
\usepackage{appendix}
\usepackage{tabularx}
\usepackage{listings}
\usepackage{color} 
\definecolor{mygreen}{RGB}{28,172,0} 
\definecolor{mylilas}{RGB}{170,55,241}

\usepackage{amsfonts}
\usepackage{amsmath}
\usepackage{amssymb}
\usepackage{bm}

\newcommand{\norm}[1]{\left|\left| #1\right|\right|}

\usepackage{cleveref}
\crefformat{section}{\S#2#1#3} 
\crefformat{subsection}{\S#2#1#3}
\crefformat{subsubsection}{\S#2#1#3}

\usepackage{subfig}
\usepackage{graphicx}
\usepackage{tikz}
\usetikzlibrary{positioning}
\usetikzlibrary{arrows.meta}
\usepackage{floatrow}
\usepackage{xspace}
\usepackage{pgfplots}
\usepackage{xcolor}

\shorttitle{A data-driven quasi-linear approximation for turbulent channel flow}
\shortauthor{J. J. Holford, M. Lee and Y. Hwang}

\title{A data-driven quasi-linear approximation for turbulent channel flow}

\author{Jacob J. Holford\aff{1\corresp{\email{jh5315@imperial.ac.uk}}}, Myoungkyu Lee\aff{2}
\and Yongyun Hwang\aff{1}}

\affiliation{\aff{1}Deparment of Aeronautics, Imperial College London, South Kensington, London, United Kingdom
\aff{2}Department of Mechanical Engineering, The University of Houston, Houston, Texas, USA}
\pgfplotsset{compat=1.17}
\begin{document}
\maketitle

\begin{abstract}
A data-driven implementation of a quasi-linear approximation is presented, extending a minimal quasi-linear approximation (MQLA) (Hwang \& Ekchardt, \textit{J. Fluid Mech.}, 2020, 894:A23) to incorporate non-zero streamwise Fourier modes. A data-based approach is proposed, matching the two-dimensional wavenumber spectra for a fixed spanwise wavenumber between a direct numerical simulation (DNS) (Lee \& Moser, \textit{J. Fluid Mech.}, 2015, 774:395-415) and that generated by the eddy viscosity-enhanced linearised Navier-Stokes equations at $Re_\tau\approx 5200$. Leveraging the self-similar nature of the energy-containing part in the DNS velocity spectra, a universal self-similar streamwise wavenumber weight is determined for the linearised fluctuation equations at $Re_\tau\simeq 5200$. This data-driven quasi-linear approximation (DQLA)  offers qualitatively similar findings to the MQLA, with quantitative improvements in the turbulence intensities and additional insights from the streamwise wavenumber spectra. By comparing the one-dimensional streamwise wavenumber spectra and two-dimensional spectra to DNS results, the limitations of the presented framework are discussed, mainly pertaining to the lack of the streak instability (or transient growth) mechanism and energy cascade from the linearised model. The DQLA is subsequently employed over a range of Reynolds numbers up to $Re_\tau = 10^5$. Overall, the turbulence statistics and spectra produced by the DQLA scale consistently with the available DNS and experimental data, with the Townsend-Perry constants displaying a mild Reynolds dependence (Hwang, Hutchins \& Marusic, \textit{J. Fluid Mech.}, 2022, 933:A8). The scaling behaviour of the turbulence intensity profiles deviates away from the classic $\ln(Re_\tau)$ scaling, following the inverse centreline velocity scaling for the higher Reynolds numbers.
\end{abstract}

\section{Introduction}
Recently, an increasingly popular approach in modelling turbulent flows is the use of the linearised Navier-Stokes equations. For example, the linearised Navier-Stokes equations are used: 1) to understand the origin of coherent structures \citep{delAlamo2006LinearChannels,Hwang2010,McKeon2010AFlow}; 2) to generate statistical inputs for the state-space estimation problem \citep{Illingworth2018EstimatingModels,Madhusudanan2019CoherentEquations,Morra2019OnFlows,Gupta2021}; 3) for a predictive quasi-linear approximation \citep{Hwang2020,Skouloudis2021ScalingApproximation}; 4) to produce a reduced order description of exact coherent states and turbulence statistics in the minimal flow unit \citep{Rosenberg2019EfficientAnalysis,Nogueira2020ForcingFlow}; 5) for statistics completion problems from partial measurements \citep{Zare2017ColourTurbulence,Towne2019}; 6) to produce reduced order models for existing flow control strategies \citep{Luhar2014OppositionFramework,Ran2021Model-basedReduction}. In wall-bounded turbulent flows, an essential feature in modelling the fluctuating state is how the nonlinear term is replaced. Since wall-bounded turbulent flows are linearly stable \citep{Butler1993OptimalFlow,Pujals2009AFlows}, a forcing or driving term is necessary to generate nontrivial solutions. Since this forcing term directly replaces the nonlinearity, how accurately this forcing statistically mimics the physics of the nonlinearity has also been observed as a key to the performance of linear models in their various uses \citep{Morra2021TheFlows,Gupta2021,Symon2023}. 

One such modelling framework, which leveraged `predictive' features of the physics of wall-bounded shear flows, was recently proposed by \cite{Hwang2020}. In this approach, referred to as minimal quasi-linear approximation (MQLA), the attached eddy model of \cite{Townsend1976} was revisited to relax the inviscid limit, allowing statistics to be predicted for high yet finite Reynolds numbers. The MQLA achieved this by following the framework of a quasi-linear approximation. The general idea of this approach is to decompose the velocity state into two separate groups, typically a large- and small-scale state. An approximation then arises when the nonlinear self-interactions of the small-scale state are neglected, and instead, a closure is provided. The earliest works implementing a quasi-linear approximation \citep{Malkus1954,malkus_1956,Herring1963,Herring1964,Herring1966} provided a closure through a marginal stability criterion. The linear stability of wall-bounded turbulence means alternative closures have to be provided. Indeed, in modern variants of quasi-linear approximations, the nonlinear self-interactions of the small-scale state have been more flexibly modelled, depending on the nature of the flow and the purpose of the approximation. Such examples include stochastic structural stability theory \citep{Farrell_Ioannou2007,farrell_ioannou_2012}, direct statistical simulation \citep{MarstonConover2008,TobiasMarston2013}, self-consistent approximations \citep{ManticLugo2014,mantič-lugo_gallaire_2016}, restricted nonlinear models \citep{ThomasEtAl2016,ThomasEtAl2015,farrell_ioannou_jiménez_constantinou_lozano-durán_nikolaidis_2016}, a quasi-linear approximation applied to exact coherent states \citep{Hall2010,Pausch_2019} and generalised quasi-linear approximations \citep{Marston2016GeneralizedJets,tobias_marston_2017,hernández_yang_hwang_2022,hernández_yang_hwang_2022_2}. The MQLA provided a closure through a stochastic forcing term, implemented self-consistently, i.e. the self-interactions of the small-scale state were enforced to be consistent with the large-scale state, in this case, the mean profile.

From the perspective of the attached eddy hypothesis, the MQLA can be regarded as a controlled approximation with the eddies arising from the linearised Navier-Stokes equations. These linear solutions replace the assumed statistical structure of the representative energy-containing eddies used by Townsend, Perry and co-workers \citep{Townsend1976,Perry1982OnTurbulence,Perry1986ATurbulence}. The solutions of the linearised Navier-Stokes equations used in the MQLA have the advantage of closely resembling a stage of the self-sustaining process \citep{Hamilton1995, Waleffe1997OnFlows}, a cycle ubiquitous in wall-bounded shear flows and the proposed mechanism for which each of the energy-containing eddies can be sustained \citep{Hwang2015}. In this linear portion of the self-sustaining process, streamwise vortices drive the formation of elongated streaks (i.e. the lift-up effect) \citep{Ellingsen1975StabilityFlow,Schmid2001StabilityFlows,Brandt2014TheFlows}, the key length scales of which are determined through the eddy viscosity enhanced linearised Navier-Stokes operator \citep{Hwang2010}. 
Instead of superposing the representative statistical structure of the energy-containing eddies subject to constant Reynolds shear stress, as done by \cite{Townsend1976}, the nonlinear term in the linearised fluctuation equations was replaced self-consistently. The linear operator was modified by an eddy viscosity diffusion term, and a forcing structure that generates the Reynolds shear stress identical to that from the nonlinear mean equation. In other words, a self-consistent closure of a quasi-linear approximation was provided. With the mean profile known, the MQLA becomes a predictive framework. The forcing term required to drive the Reynolds shear stress consistent with the mean profile allows further statistics of the velocity fluctuations to be determined at different Reynolds numbers.

While the attached eddy hypothesis assumes and leverages the self-similar nature of the eddies, attention has been turned to understanding the statistical structure of the forcing or nonlinear term. The benefits of having a well-prescribed model for the nonlinear term are demonstrated, for example, by the performance of state-space estimators \citep{hœpffner_chevalier_bewley_henningson_2005,Chevalier2006StateFlows,Illingworth2018EstimatingModels,Madhusudanan2019CoherentEquations,Illingworth2018EstimatingModels,Morra2019OnFlows}, with more physical models for the nonlinear term resulting in better predictions of velocity statistics across the wall-normal direction \citep{Gupta2021}. There are a variety of approaches in determining and modelling the forcing term, ranging from statistically exact control and optimisation-based techniques \cite[for a review, see ][]{Jaovanovic2021} to measurement through direct numerical simulation \citep{Chevalier2006StateFlows,Nogueira2020ForcingFlow,Morra2021TheFlows}, as well as more phenomenological modelling \citep{Jovanovic2001, Gupta2021, Holford2022}. For example, \cite{Zare2017ColourTurbulence} determined a set of two-point coloured-in-time forcing statistics through an optimisation problem. This optimisation problem had constraints such that the velocity spectral density exactly matched that of a direct numerical simulation (DNS). This work was recently complemented by \cite{abootorabi_zare_2023}, demonstrating the benefits of an additional eddy-viscosity diffusion operator to the coloured-in-time forcing. While these methods can yield forcing statistics that generate the exact velocity statistics, often, the techniques are local in the sense that they are applied to a single wavenumber pair and hence have to be repeated for every wavenumber pair to build a global forcing structure. 

Regarding implementing a predictive framework, the inputs to many models are often obtained through DNS, which is frequently a desirable output. To address this issue, \cite{Holford2022} recently identified a global forcing structure across the entire wavenumbers required for the eddy-viscosity enhanced linearised Navier-Stokes operator with a simplification that the forcing is white in time and decorrelated in space. The findings revealed the self-similar nature of the forcing spectra corresponding to the main energy-containing part of the velocity spectra. In the current work, \cite{Holford2022} is approximately followed to provide a global structure for the forcing across streamwise wavenumbers and Reynolds numbers. 


The main findings of the MQLA \citep{Hwang2020} were consistent with the seminal predictions of Townsend's attached eddy hypothesis on turbulence intensities. By superposing the solutions of the linearised Navier-Stokes equations with a forcing that provides a self-consistent Reynolds shear stress, a logarithmic decay in the wall-parallel turbulence intensities was found, as well as a region where the wall-normal turbulence intensity is approximately constant. Additionally, since the determination of the fluctuating velocity generated Reynolds shear stress in the MQLA requires integration of the velocity spectra, the scaling behaviour of the one-dimensional velocity spectra could also be extrapolated to exceptionally high Reynolds numbers \citep{Skouloudis2021ScalingApproximation}. A strong qualitative match was found between the one-dimensional spanwise wavenumber velocity spectra produced by the MQLA and those reported by direct numerical simulation (DNS) \citep{Lee2015}. However, important quantitative differences were also found. In particular, turbulence intensities of the MQLA were highly anisotropic compared to those of DNS, with the streamwise turbulence intensity far exceeding that of DNS. At the same time, the other velocity components, particularly the wall-normal component, were significantly lower. This result stems from neglecting the streamwise varying Fourier modes in the MQLA, causing the absence of streamwise pressure strain that transfers the energy produced in the streamwise component to the other components \citep{Cho2018ScaleFlow,Lee2019SpectralNumber}. Neglecting the streamwise varying Fourier modes is also understood to prohibit the MQLA's capabilities in reproducing features of the self-sustaining process beyond the lift-up effect, particularly the streak instability or transient growth, which play a crucial role in redistributing turbulent kinetic energy from the streamwise velocity component to the others \citep{Schoppa2002CoherentTurbulence,deGiovanetti2017StreakMotions,Doohan2021MinimalTurbulence,Lozano-Duran2021Cause-and-effectTurbulence}.

In light of how neglecting the streamwise varying Fourier modes leads to highly anisotropic results, the present paper aims to extend the MQLA to include the streamwise varying Fourier modes, with an emphasis on maintaining the predictive nature of the MQLA at different Reynolds numbers. In the MQLA, limiting the framework to consider only the streamwise uniform Fourier modes leaves mainly the lift-up mechanism in generating the turbulence statistics \citep{Jiao2021OrrFlow}. By relaxing this two-dimensional assumption, continuity reduces the anisotropy towards the streamwise turbulence statistics. However, incorporating the streamwise varying Fourier modes into the present quasi-linear framework is still unable to account for the streak instability and/or transient growth mechanism \citep{Schoppa2002CoherentTurbulence,deGiovanetti2017StreakMotions,Lozano-Duran2021Cause-and-effectTurbulence}, which is also understood to be involved in the determination of the streamwise length scale of energy-containing eddies. In order to overcome these difficulties, a physics-aware data-driven approach will be taken. Hereafter, the framework of the current study will be referred to as the data-driven quasi-linear approximation (DQLA) to distinguish it from the MQLA. The DQLA framework incorporates the physics element through the attached eddy hypothesis
and enforces self-similarity of the forcing with respect to the spanwise length scale in the streamwise direction. The detailed structure of the forcing is determined using the DNS database from \citet{Lee2015}. This approach relies on the self-similar nature of the eddy viscosity-enhanced linear operator \citep{Hwang2010}, which has been corroborated by recent findings on the statistical structure of the forcing of this linear model from \citet{Holford2022}.

The paper is organised as follows. The DQLA framework is developed in \S\ref{sec:ProblemFormulation}, formulating two optimisation problems. Firstly a self-similar weight is determined following \cite{Holford2022}. This weight is then extrapolated across all considered Fourier modes, and a quasi-linear approximation is then implemented following \cite{Hwang2020}. The linear model used throughout this study includes an eddy-viscosity diffusion operator, and its significance is briefly discussed. 
The results of the DQLA at $Re_\tau = 5200$ are then compared with statistics from DNS of \cite{Lee2015} in \S\ref{sec:DDQLAre5200}. Emphasis will be placed on the streamwise one-dimensional and two-dimensional velocity spectra since the MQLA cannot produce these statistics. Additionally, any quantitative improvements of the DQLA compared to the MQLA will be discussed concerning the inclusion of streamwise varying Fourier modes. The predictive capabilities of the DQLA will be assessed by extrapolating results up to $Re_\tau = 10^5$ in \S\ref{sec:DDQLAscaling}. The paper is then concluded in section \ref{sec:conclusions}, which summarises the results and limitations of this framework. 

\section{Problem formulation}\label{sec:ProblemFormulation}
\subsection{Turbulent channel flow}
A quasi-linear approximation for incompressible, fully-developed turbulent channel flow is considered. The flow domain is the region confined between two infinitely long and wide plates. The coordinates along the streamwise, wall-normal and spanwise directions are denoted by $x$, $y$ and $z$, respectively, with a corresponding velocity vector $\bm u = \left(u,\,v,\,w\right)$. The two plates are located at $y=0,2h$, where $h$ represents the half-width of the channel. The velocity field is decomposed into a time-averaged mean-field, $\bm U = \left(U(y),\,0,\,0\right)$, and the fluctuating velocity about this mean profile, $\bm u' = \left(u',\,v',\,w'\right)$, i.e. the Reynolds decomposition. This results in the following coupled set of equations:
\begin{subequations}
    \begin{equation}\label{subeq:meanEquation}
    \nu\frac{\mathrm{d}U}{\mathrm{d}y} - \overline{u'v'} = \frac{\tau_w}{\rho}\left(1-\frac{y}{h}\right),
    \end{equation}
    \begin{equation}\label{subeq:fluctuatingVelocity}
    \frac{\partial \bm u'}{\partial t} + \left(\bm U \cdot \bm \nabla\right)\bm u' + \left(\bm u' \cdot \bm \nabla\right)\bm U = -\frac{1}{\rho}\bm \nabla p + \nu \nabla^2\bm u + \mathcal{\bm N},
    \end{equation}
    where
    \begin{equation}\label{subeq:exactNLterm}
        \mathcal{N} = -\bm \nabla \cdot\left(\bm u'\bm u' - \overline{\bm u' \bm u'}\right).
    \end{equation}
Here, $t$ denotes time, $\rho$ the fluid density,  $p’$ the fluctuating pressure, $\nu$ the kinematic viscosity, $\overline{(\,\,\cdot\,\,)}$ a time-averaged quantity and $\tau_w$ is the time averaged wall shear stress.
\end{subequations} Following the typical quasi-linear approximation framework, the time-averaged mean profile equation retains this form, including the nonlinear Reynolds shear stress term feeding back from the fluctuating velocity field.  The dynamics of the fluctuating velocity are then `linearised', dropping the self-advection term. As the associated linear operator is stable for the turbulent mean profile in channel flow \citep{Pujals2009AFlows}, an additional driving term must be considered for nontrivial statistics. To this end, the nonlinear term is replaced with the following model:
\begin{subequations}
    \begin{equation}\label{subeq:modelNLterm}
        \mathcal{N}_{\nu_t,f} = \bm \nabla \cdot \left(\nu_t (\bm \nabla \bm u' + \bm \nabla \bm u'^T)\right) + \bm f',
    \end{equation}
    where $\bm f'=(f_u',f_v',f_w')$ is a stochastic forcing term and Cess' expression \citep{Cess1958AFlow} is used for the eddy viscosity profile,
    \begin{equation}\label{subeq:ReynoldsEddyVisc}
        \nu_t\left(\eta\right) = \frac{\nu}{2}\left\{1 + \frac{\kappa^2Re_{\tau}^2}{9}\left(1 - \eta^2\right)^2\left(1+2\eta^2\right)^2\left(1-\exp\left[\left(\left|\eta\right|-1\right)Re_{\tau}/A\right]\right)^2\right\} - \frac{\nu}{2}
    \end{equation}
    with $\eta = \left(y-h\right)/h$. 
\end{subequations}
Including the eddy viscosity term in (\ref{subeq:modelNLterm}) is not a necessity. Using only a forcing term would leave the coupled system as statistically `exact' if the forcing was set to be identical to a set of known statistics generated by the nonlinear term. The eddy viscosity is used as it provides a simple crude approximation for the role of the nonlinear term, removing energy across all considered integral length scales and Reynolds numbers \citep{Hwang2020, Symon2021EnergyModelling}. The same eddy viscosity is also used as a closure in determining the mean velocity profile with 
\begin{equation}\label{eq:meanUV}
    -\overline{u'v'} = \nu_t\frac{\mathrm{d}U}{\mathrm{d}y},
\end{equation}
upon which solving \eqref{subeq:meanEquation} gives the robust law of the wall. Following this mean profile closure, the eddy viscosity parameters are set with $\kappa = 0.426$ and $A = 25.4$, obtained by the best least squares fitting the mean profile obtained by integrating \eqref{subeq:meanEquation} with \eqref{eq:meanUV} and the DNS mean profile at $Re_\tau \approx 2000$ \citep{delAlamo2006LinearChannels}. 
The crude physical argument for using both the same eddy viscosity profile in the mean and fluctuating velocity component equates to both velocity fields experiencing the same background turbulence. While this physical assumption is mainly used out of simplicity, the inclusion of an eddy viscosity diffusion operator has been shown to be beneficial in many previous studies \cite[see, for example,][]{Pujals2009AFlows,Zare2017ColourTurbulence,Illingworth2018EstimatingModels,Morra2019OnFlows,Symon2021EnergyModelling}. Notably, the inclusion of the eddy viscosity-based diffusion enables one to describe the inner-scaling behaviour of the near-wall attached region of the outer scale structures observed in full DNS, in terms of the modes associated with transient growth, resolvent analysis and the stochastic response of the linearised fluctuation equations \citep{Hwang2010,Hwang2016MesolayerFlow}.


In the present study, the forcing is first considered to be white in time and decorrelated in the wall-normal direction for the purpose of utilising the framework of the stochastic linear dynamical system \citep{Farrell1992StochasticEquations,Jovanovic2005ComponentwiseFlows,Hwang2010}. Given the homogeneous nature in the wall-parallel directions, it is convenient to consider the Fourier transform along those directions:
\begin{equation}\label{eq:FourierTransform}
    \hat{\bm f}'(t,y;k_x,k_z) = \int_{-\infty}^{\infty}\int_{-\infty}^{\infty} \bm f'(t,x,y,z)e^{i(k_x x + k_z z)}\mathrm{d}x \mathrm{d}z,
\end{equation}
giving the wall-normal forcing profiles at a given pair of streamwise and spanwise length scales $\lambda_x = {2\pi}/{k_x}$ and $\lambda_z = {2\pi}/{k_z}$, where ($k_x, k_z$) is the considered wavenumber pair. Analogous definitions of the Fourier transform are used for the other flow states. The spectral covariance matrix for the forcing is then considered to be
\begin{equation}\label{eq:forcingcov}
    \mathbb{E}\left[\hat{\bm f}'(y,t;k_x,k_z)\hat{\bm f}'^H(y',t';k_x,k_z) \right] =  \begin{bmatrix}
        W_u(k_x,k_z) & 0  & 0 \\ \,\, 0 & W_v(k_x,k_z) & 0\\ \,\,0& 0& W_w(k_x,k_z)
    \end{bmatrix}\delta(y-y')\delta(t-t'),
\end{equation}
where $(\cdot)^{H}$ denotes complex conjugate transpose, $\mathbb{E}\left[\cdot \right]$ is the expectation operator over different stochastic realisations, and $W_{r}$, with $r=\{u,v,w\}$, are componentwise weights to be determined. Here, the forcing amplitude is considered to vary componentwise, and this is to model the anisotropic nature of the velocity statistics and spectra in channel flow more flexibly.

The resulting power- and cross-spectral densities of velocity fluctuations are obtained from the following velocity spectral covariance matrix:
\begin{subequations}
\begin{equation}
    \Phi_{\bm u \bm u}(y,y';k_x,k_z) = \mathbb{E}\left[\hat{\bm u}'(y,t;k_x,k_z)\hat{\bm u}'^H(y',t;k_x,k_z)\right].                                  
\end{equation}
Given the linear relation between the velocity and forcing spectral covariance matrices \citep{Farrell1992StochasticEquations,Jovanovic2005ComponentwiseFlows,Hwang2010}, $\Phi_{\bm u \bm u}(y,y';k_x,k_z)$ can further be decomposed into
\begin{eqnarray}
    \Phi_{\bm u \bm u}(y,y';k_x,k_z) = \sum_{r=u,v,w}W_{r}(k_x,k_z)\Phi_{\bm u \bm u,r}(y,y';k_x,k_z), \nonumber
\end{eqnarray}
\end{subequations}
where $\Phi_{\bm u \bm u,r}(y,y';k_x,k_z)$ is the velocity spectral covariance matrix associated with each component of the forcing with the unit amplitude: for example, $\Phi_{\bm u \bm u,u}$ is obtained by setting the forcing spectral covariance to be
\begin{equation}\label{eq:whiteForceU}
    \mathbb{E}\left[\hat{\bm f}'(y,t;k_x,k_z)\hat{\bm f}'^H(y',t';k_x,k_z) \right] = \begin{bmatrix}
        \delta(y-y') &  0 &  0\\ \,\, 0 & 0 & 0\\ \,\,0& 0& 0
    \end{bmatrix}\delta(t-t'),
\end{equation}
and $\Phi_{\bm u \bm u,v}$ and $\Phi_{\bm u \bm u,w}$ are obtained in the same manner. In this study, $\Phi_{\bm u \bm u,r}$ for $r=\{u,v,w\}$ are obtained by solving the standard Lyapunov equation formulated with the Orr-Sommerfeld-Squire system of equations. The equations are discretized using a Chebyshev collocation method \citep{Weideman2000ASuite} with the wall-normal grid points reported in Table \ref{tab:errorsTable}. For the details of the solution method, the reader may refer to previous studies \citep{Hwang2010,Holford2022}. The domain of streamwise and spanwise wavelengths are considered as $(\lambda_x,\lambda_z) \in [10\delta_\nu,10\delta_\nu] \times [100h,10h]$ ($\delta_\nu=\nu/u_\tau$, where $u_\tau$ is the friction velocity) to cover a range of length scales including near-wall motions, as well as very large-scale motions in the outer region.

\subsection{Simplifications}
The forcing covariance considered in (\ref{eq:forcingcov}) may be a starting point of the proposed quasi-linear approximation. However, a white-in-time, uniform forcing in the wall-normal direction at each $k_x$ and $k_z$ is non-physical, as discussed in detail in \cite{Holford2022}. Furthermore, applying such a forcing in the MQLA yielded erroneous behaviours for the spectra associated with energy cascade \cite[]{Hwang2020}. It was previously shown that considering some of the leading POD modes from the response to this forcing offers an effective means to filter out the non-physical part in the resulting velocity spectra while retaining the physically relevant part that models the energy-containing motions at integral length scales \citep{Hwang2020}. Following this approach, the forcing statistics in this study are also implicitly modified to drive leading POD modes that contain significant energetic content of the overall response \citep{Hwang2010}. Consequently, for a given wavenumber pair, the velocity spectral covariance matrix for each forcing component (i.e. $\Phi_{\bm u \bm u,r}$) is further approximated in terms of the leading order POD modes, such that
\begin{subequations}
\begin{equation}\label{subeq:PODreconstruction}
    \Phi_{\bm u \bm u,r}^{N_{\mathrm{POD}}}(y,y';k_x,k_z)=\sum_{i=1}^{N_{\mathrm{POD}}}\sigma_{i} \hat{\bm u}_{i,r,\mathrm{POD}}(y;k_x,k_z)\hat{\bm u}^H_{i,r,\mathrm{POD}}(y';k_x,k_z),
\end{equation}
where $N_{\mathrm{POD}}$ is the number of POD modes, and $\sigma_{i}$ and $\hat{\bm u}_{i,r,\mathrm{POD}}(y';k_x,k_z)$ are the eigenvalues and eigenfunctions of the original velocity spectral covariance matrix with white-in-time and spatially-decorrelated forcing, denoted by $\Phi_{\bm u \bm u}^W$:
\begin{equation}{\label{eq:PODreconstructVel}}
   \int_0^{2h} \Phi_{\bm u \bm u}^{W}(y,y';k_x,k_z) \hat{\bm u}_{i,\mathrm{POD}}(y';k_x,k_z)\textrm{d}y' = \sigma_i \hat{\bm u}_{i,\mathrm{POD}}(y;k_x,k_z)
\end{equation}
with $\sigma_i \geq \sigma_{i+1}.$
\end{subequations} 
Considering the previous observation in \cite{Hwang2020}, $N_{\mathrm{POD}} = 2$ is chosen here, retaining the most energetic structure when accounting for the geometrical symmetry in channel flow about $y=h$. 

Further to this, the weighting along the streamwise wavenumber axis is implemented considering the self-similarity of the forcing structure with respect to the spanwise wavenumber \citep{Hwang2010,Holford2022}, as expected from the attached eddy hypothesis of Townsend \cite[]{Townsend1976,Hwang2015} (see \S\ref{sec:IIC} for further details). 
Consequently, the weight of each component is decomposed into a part retaining the self-similar structure $W_{r,k_x}(k_x/k_z)$ along the streamwise wavenumber axis and a part determining its amplitude $W_{k_z}(k_z)$ for each spanwise wavenumber, such that
\begin{equation}
    W_{r}(k_x,k_z) = W_{r,k_x}(k_x/k_z)W_{k_z}(k_z),
\end{equation}
giving the final form of velocity spectra covariance matrix for the quasi-linear approximation in this study:
\begin{equation}{\label{eq:reconstructVeloictyFinal}}
    \Phi_{\bm u \bm u}(y,y';k_x,k_z) = W_{k_z}(k_z)\sum_{r=u,v,w}W_{r,k_x}(k_x/k_z)\Phi_{\bm u \bm u,r}^{N_\mathrm{POD}}(y,y';k_x,k_z).
\end{equation}
The weights are determined by solving the optimisation problems proposed in the two following subsections. 

\subsection{Data-driven determination of streamwise weighting}\label{sec:IIC}

To first determine the self-similar weight $W_{r,k_x}(k_x/k_z)$ along the streamwise wavenumber axis, for each $k_z$, an optimisation problem is considered such that \eqref{eq:reconstructVeloictyFinal} best matches the two-dimensional velocity spectra and Reynolds shear stress cospectra of DNS at $Re_\tau \approx 5200$ \citep{Lee2015}, denoted by $\Phi_{\bm u \bm u }^{\mathrm{DNS}}(y;k_x,k_z)$. This problem has recently been solved with the addition of a wall-normal variation in the forcing term in \cite{Holford2022}, and the reader is referred to it for a more complete discussion on the modelling rationale. The present study leverages the main findings from \cite{Holford2022}. However, a more straightforward problem is solved for the weight, with the use of the POD modes implicitly varying the wall-normal profile of the forcing term so that the modelling efforts can be extrapolated to other Reynolds numbers without using any further DNS data at different Reynolds numbers.

To determine the weight, two main observations are highlighted. The first is the self-similar nature of the velocity spectra from DNS with respect to the spanwise length scale \citep{Hwang2015,Holford2022}. In particular, the self-similarity occurs at wavenumbers which contribute significantly to the turbulence intensity profiles, i.e. main energy-containing features of the spectra. The second is that the linearised Navier-Stokes equations also generate an approximately self-similar response with respect to the spanwise length scale \citep{Hwang2010, Hwang2020}. Combining these two observations, an optimisation problem that weights a self-similar linear response to an approximately self-similar set of DNS spectra is considered. It is then expected that the weights themselves would be self-similar, at least to the same degree as the DNS velocity spectra. A weighting is now determined to match the two-dimensional velocity spectra from DNS at a given spanwise length scale through the following optimisation problem:
\begin{subequations}\label{eq:entireOptizationProblem}
\begin{equation}\label{eq:NormMin}
    \min_{W_{r,k_x}}\sum_s\frac{\norm{\bm \Phi^{\mathrm{DNS}}_{s} - \bm \Phi_{s}(W_{r,k_x})}_Q}{\norm{\bm \Phi^{\mathrm{DNS}}_{s}}_Q} +\sum_{r}\gamma{J[W_{r,k_x}]},
\end{equation}
subject to 
\begin{equation}
    W_{r,k_x}(k_x,y;k_z) \geq 0,
\end{equation}
where $r = \{u, v, w\}$, $s = \{uu, vv, ww, uv\}$ and $\gamma$ is a parameter controlling the relative importance of the regularisation. Here, $J$ is a regularisation functional to penalise the smoothness of the forcing intensity profiles, chosen to be
\begin{equation}
    J[W_{r,k_x}] = \norm{\left(\frac{\partial^2 W_{r,k_x}}{\partial \ln k_x\,^2}\right)}_Q,
\end{equation}
\end{subequations}
where $\norm{\cdot}_Q$ is a norm defined as $\norm{\cdot}_Q^2=\int_0^{2h}\int_{-\infty}^{\infty}\left(\cdot\right)^2k_xQ(y)\mathrm{d}\ln k_x\mathrm{d}y$ with weight $Q(y) = \chi ^{-1}$, where $\chi = 1 - |\eta|$ is the distance from the wall, to place equal emphasis on points following a logarithmic scaling with distance from the wall. Note that the logarithmic coordinates are also used along the $k_x$-axis to focus the problem on the modelling of the self-similar energy-containing part, placing less significance on the non-self-similar-part originating from energy cascade \cite[see][for a further discussion]{Holford2022}. In particular, the purpose of this weighting along the streamwise wavenumber axis is to provide a self-similar weight for use across all wavenumber pairs and all Reynolds numbers.  

The convex optimisation problem \eqref{eq:entireOptizationProblem} was solved by discretising the spectral velocity state onto $N_y = 512$ grid points with the Chebyshev-collocation method. Discretisation along the streamwise wavenumber axis was carried out with logarithmic spacing to maintain ${\Delta} (\ln k_x) \leq 0.05$ or otherwise to align with the DNS streamwise wavenumbers for the smaller values of $k_xh$, resulting in 132 streamwise Fourier modes being used, and integration performed using the trapezoidal rule along the streamwise wavenumber axis. The optimisation problem was then converted to a standard second-order cone problem and solved with the MOSEK solver \citep{mosek}. The trend in the optimisation errors, defined by the first term in (\ref{eq:NormMin}), upon increasing the regularisation parameter, was found to be monotonically increasing with $\gamma$ (see figure \ref{fig:gammaTrace} in appendix \ref{sec:sensOfWeight}). Hence, $\gamma$ was set upon inspection of the velocity spectra and smoothness of the weights. An approximate value of $\gamma = 0.5$ was used and changed upon trial and inspection, giving values ranging from 0.3-1.2 across the considered spanwise wavenumbers. 

\begin{figure}
    \begin{center}
    \sidesubfloat[]{\hspace{-0.6cm}
        \begin{tikzpicture}
            \node (img)  {\includegraphics[width = 4.6cm]{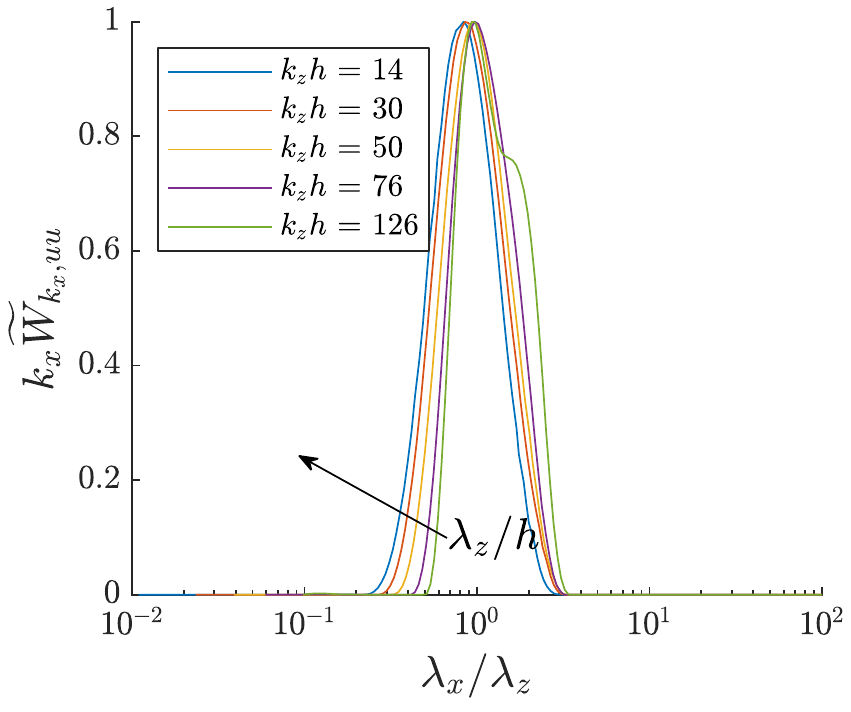}};
        \end{tikzpicture}
        \label{subfig:WuuSelfSim}}\hspace{-0.8cm}
    \sidesubfloat[]{\hspace{-0.6cm}
        \begin{tikzpicture}
          \node (img)  {\includegraphics[width = 4.6cm]{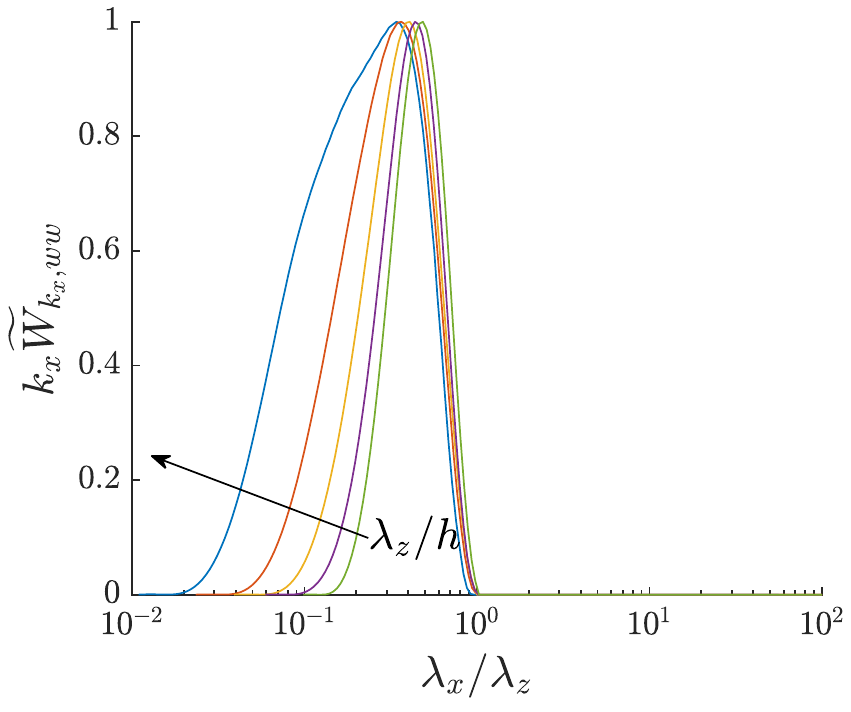}};
        \end{tikzpicture}
        \label{subfig:WwwSelfSim}} \hspace{-0.8cm} 
    \sidesubfloat[]{\hspace{-0.6cm}
        \begin{tikzpicture}
            \node (img)  {\includegraphics[width = 4.6cm]{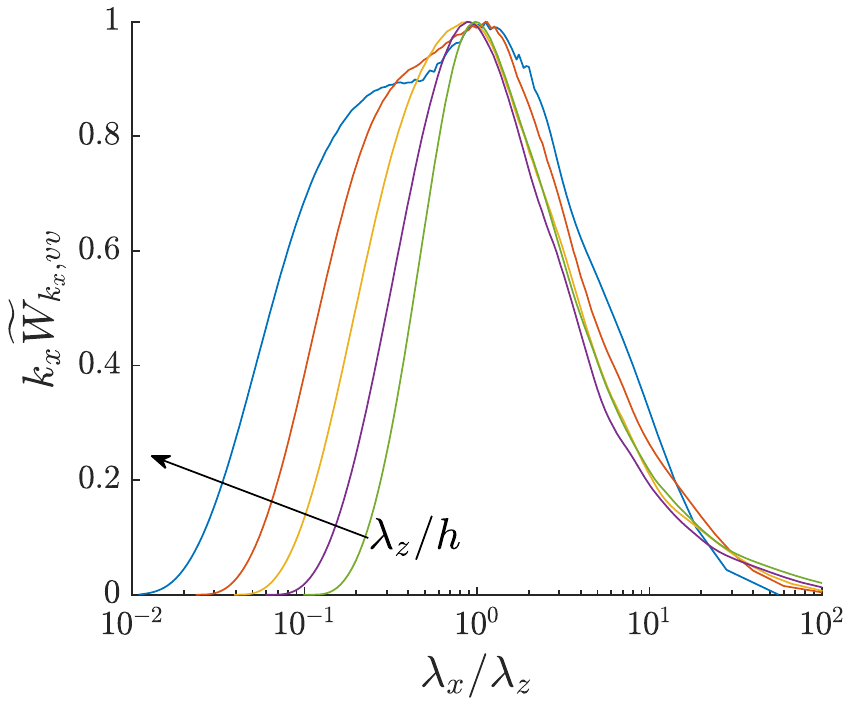}};
        \end{tikzpicture}
        \label{subfig:WvvSelfSim}}
    \end{center}
    \caption{Premultiplied streamwise Fourier mode weights in self-similar coordinates for $k_zh =  14, 30, 50, 76, 126$ ($\lambda_z/h=0.45, 0.21, 0.13, 0.08, 0.05$ or $\lambda_z^+= 2327, 1086, 651, 429, 259$): (a) streamwise, (b) spanwise and (c) wall-normal velocity components. Here the $\widetilde{(\cdot)}$ denotes the weights normalised such that their premultiplied maximum value is one. }\label{fig:selfSimWeighting}
\end{figure}

\begin{figure}
    \begin{center}
    \sidesubfloat[]{\hspace{-1.1cm}
        \begin{tikzpicture}
            \node (img)  {\includegraphics[width = 5.0cm]{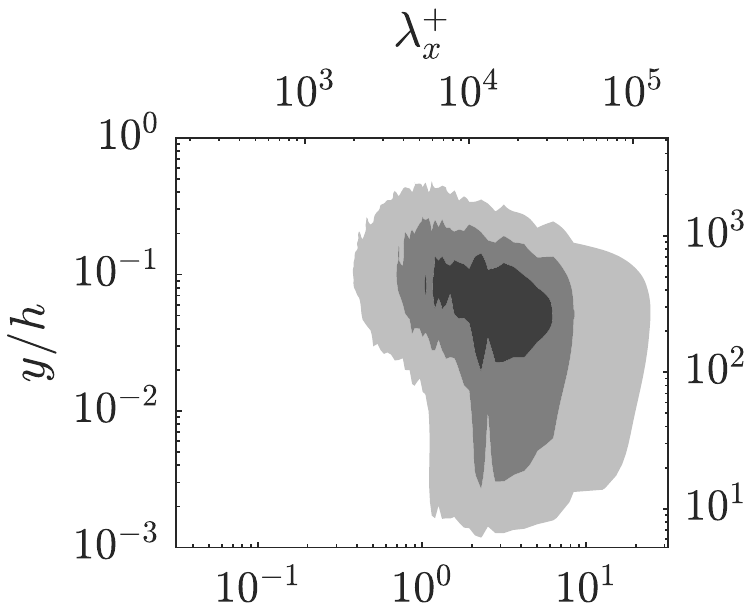}};
        \end{tikzpicture}
        \label{subfig:PhiuuKzh14DNS}}\hspace{1cm}
    \sidesubfloat[]{\hspace{-0.5cm}
        \begin{tikzpicture}
            \node (img)  {\includegraphics[width = 5.0cm]{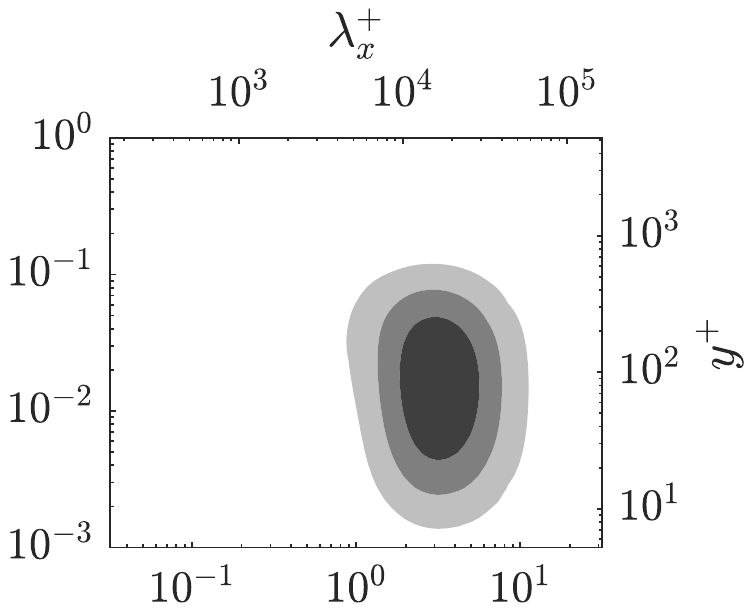}};
        \end{tikzpicture}
        \label{subfig:PhiuuKzh14QLA}} \hfill \\
    \sidesubfloat[]{\hspace{-1.1cm}
        \begin{tikzpicture}
            \node (img)  {\includegraphics[width = 5.0cm]{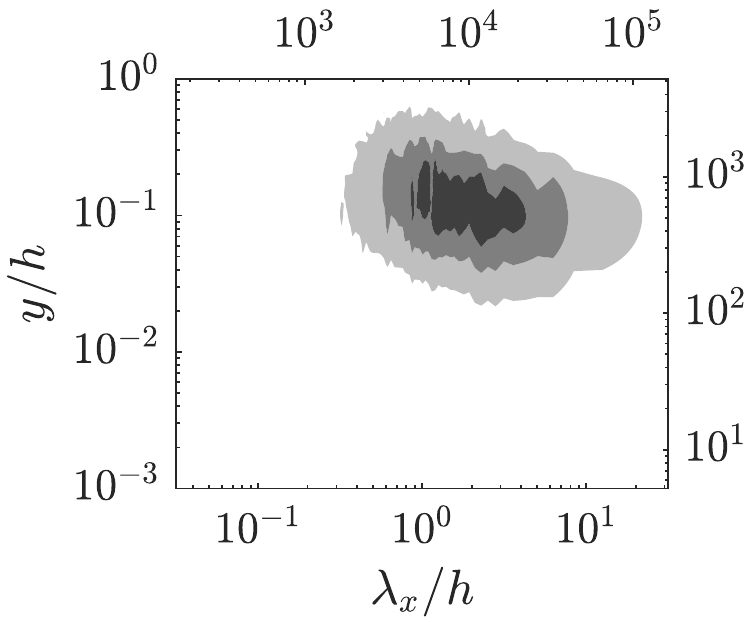}};
        \end{tikzpicture}
        \label{subfig:PhiuvKzh14DNS}}\hspace{1cm}
    \sidesubfloat[]{\hspace{-0.5cm}
        \begin{tikzpicture}
            \node (img)  {\includegraphics[width = 5.0cm]{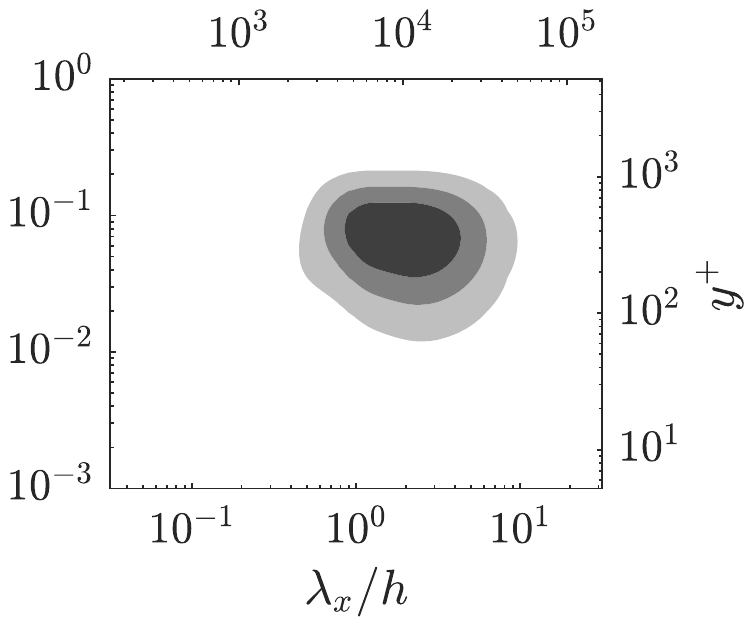}};
        \end{tikzpicture}
        \label{subfig:PhiuvKzh14QLA}}
    \end{center}
    \caption{Premultiplied two-dimensional spectra in self-similar coordinates for $k_zh = 14$, $k_z^+ = 0.0027$ ($\lambda_z = 0.45h$, $\lambda_z^+ = 2333$) of (a,b) streamwise velocity spectra and (c,d) Reynolds shear stress cospectra determined from (a,c) DNS and (b,d) DQLA. The contour levels are separated by 0.25 times the maximum value for each spectra.}\label{fig:kzh14Spectra}
\end{figure}

Figure \ref{fig:selfSimWeighting} shows solutions to \eqref{eq:entireOptizationProblem} for spanwise length scales associated with the logarithmic region, varying from $k_zh = 14$ ($\lambda_z/h \approx 0.45$) up to  $k_zh = 126$ ($\lambda_z^+ \approx 259$). 
 Here, the premultiplied weighting is plotted, as this is more physically representative of the forcing spectral density in logarithmic coordinates, as opposed to a weighting correction applied to the velocity spectra. A similar qualitative trend is seen across all components: an approximate self-similarity at relatively large wavelengths ($\lambda_x \gtrsim \lambda_z$) in the weighting, with a spanwise-wavelength-dependent weighting at small wavelengths  ($\lambda_x \lesssim \lambda_z$). This trend qualitatively agrees well with the recent finding in \cite{Holford2022}, where the forcing spectra, obtained for the same linear model, were found to be approximately self-similar at the integral length scale (i.e. $\lambda_x \sim \lambda_z \sim y$), while they are not for $\lambda_x \lesssim \lambda_z$ at which the corresponding DNS spectra are associated with energy cascade. In particular, the forcing spectra for $\lambda_x \lesssim \lambda_z$ were also found to grow with $\lambda_z$ due to the increased separation of local integral and dissipation length scales, consistent with the behaviours of the weights in figure \ref{fig:selfSimWeighting}. 
That being said, it is important to note that this scale-dependent weighting for $\lambda_x \lesssim \lambda_z$ has little significance in modelling the spectra \citep{Holford2022}. 
At the smaller streamwise length scales of $\lambda_x \lesssim \lambda_z$, large forcing input is required to drive relatively inconsequential features of the velocity spectra associated with the energy cascade \citep{Holford2022} (see the discussion below and figure \ref{subfig:PhiuuKzh14QLA}). The contributions of these features to the turbulent kinetic energy are small, and they have often been ignored in the classical attached eddy models \citep{Townsend1976,hwang_hutchins_marusic_2022}. In a similar context, no attempt is made here to correct or modify the weighting at different length scales. This observation is further confirmed by assuming the weights given from solutions of \eqref{eq:entireOptizationProblem} at different $\lambda_z$ as self-similar and determining the errors between the normalised spectra at different spanwise length scales (see figure \ref{subfig:WkxSensTwoDspec} in Appendix \ref{sec:sensOfWeight} for the sensitivity to weights). It is shown that  the scale-dependent features in the weight of $W_{r,k_x}(k_x/k_z)$ have only a small effect on the structure of the velocity spectra, with the total errors being largely independent of the weight obtained for different $k_z$ in figure \ref{fig:selfSimWeighting}.

To demonstrate what kind of two-dimensional velocity spectra in the $k_x$-$y$ plane \eqref{eq:entireOptizationProblem} determines, figure \ref{fig:kzh14Spectra} compares the streamwise velocity spectra (figures \ref{subfig:PhiuuKzh14DNS} and \ref{subfig:PhiuuKzh14QLA}) and Reynolds shear-stress cospectra (figures \ref{subfig:PhiuvKzh14DNS} and \ref{subfig:PhiuvKzh14QLA}) of the DNS, to the optimally weighted velocity spectra at $k_zh = 14$. In the streamwise component of the optimally weighted spectra (figure \ref{subfig:PhiuuKzh14QLA}), an energetic response is seen for $1 \lesssim \lambda_x/h \lesssim 10$, consistent with the DNS streamwise velocity spectra (figure \ref{subfig:PhiuuKzh14DNS}). However, the DNS streamwise velocity spectra also have energetic content for $\lambda_x/h \lesssim 1$ away from the wall ($y/h \gtrsim 0.02$). This part of the spectra has been understood to be associated with the physical processes that would not be modelled well by the linearised model (\ref{subeq:fluctuatingVelocity}) with (\ref{subeq:modelNLterm}), as discussed in detail in the previous studies \citep{Hwang2015,deGiovanetti2017StreakMotions,Holford2022}: i.e. streak instability or transient growth in self-sustaining process \citep{Hamilton1995,Schoppa2002CoherentTurbulence,Lozano-Duran2021Cause-and-effectTurbulence}, and the resulting energy cascade and dissipation \citep{Doohan2021MinimalTurbulence}. The weighted linear response of leading POD modes appears to neglect a majority of these features. 
The other qualitative difference between the spectra is the extent to which the primary peak extends towards the wall, with the primary peak from the linear response extending close towards the wall. The observed differences between the DNS streamwise velocity spectra and the optimally weighted leading POD modes indicate some critical limitations in the current modelling approach. It may be overcome by taking the recent approaches designed to fully reconstruct the spectra using the given linear model at a given Reynolds number \citep{abootorabi_zare_2023}. However, it is worth mentioning that such approaches do not easily enable us to use the modelled spectra for extrapolation to other Reynolds numbers without additional DNS data. For this purpose, a simple approach is taken, and the spectra not directly associated with the linear processes of the flow are ignored. As discussed above, this is similar to the original attached eddy model of Townsend \citep{Townsend1976}, which ignored all the motions related to energy cascade and dissipation (i.e. small-scale detached eddies). 


A more substantial qualitative match is found comparing the Reynolds shear stress cospectra (figures \ref{subfig:PhiuvKzh14DNS} and \ref{subfig:PhiuvKzh14QLA}). 
The weighted linear response provides a good agreement for the streamwise wavelength of the primary peak and is overall energetic for a majority of the corresponding DNS spectra. However, it should be noted that the overall amplitude of the Reynolds shear stress generated by the optimally weighted leading POD modes is relatively weak. 
For the standard $L^2$-norm over the entire wall-normal distance and considered streamwise wavenumbers (i.e. the root mean square velocities), the ratio of the norms of Reynolds shear stress between the linear model and DNS is approximately $0.4$. In contrast, the streamwise response is approximately $0.9$, indicating a relatively low Reynolds shear stress. 
This level of anisotropy is expected to carry through results, an intrinsic limitation in this modelling approach, and it originates from the spatially decorrelated nature of the forcing considered initially \cite[see also][for a further discussion]{Holford2022}. Given that the weight shown in figure \ref{fig:selfSimWeighting} is approximately self-similar for $\lambda_x/\lambda_z \gtrsim 1$ and the non-self-similar parts for $\lambda_x/\lambda_z \lesssim 1$ do not generate a strong response for the resulting spectra in figure \ref{fig:kzh14Spectra}, the weight from $k_zh = 30$ shall be used as a self-similar weight for all wavenumber pairs throughout this study. Note that the choice of the weight does not significantly change the resulting quasi-linear approximation (see Appendix \ref{sec:sensOfWeight} for the sensitivity to weight choice on results of the entire DQLA procedure at $Re_\tau \approx 5200$). 

\begin{figure}
    \begin{center}
    \sidesubfloat[]{\hspace{-0.5cm}
        \begin{tikzpicture}
            \node (img)  {\includegraphics[width = 5.0cm]{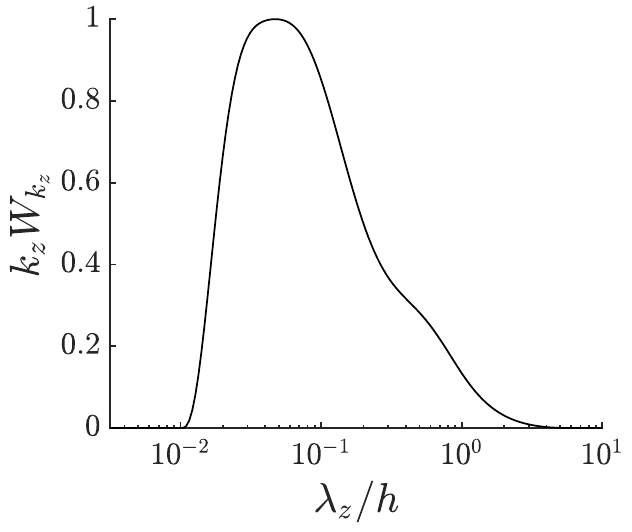}};
        \end{tikzpicture}
        \label{subfig:SpanWeight}}\hspace{1cm}
    \sidesubfloat[]{\hspace{-0.5cm}
        \begin{tikzpicture}
          \node (img)  {\includegraphics[width = 5.0cm]{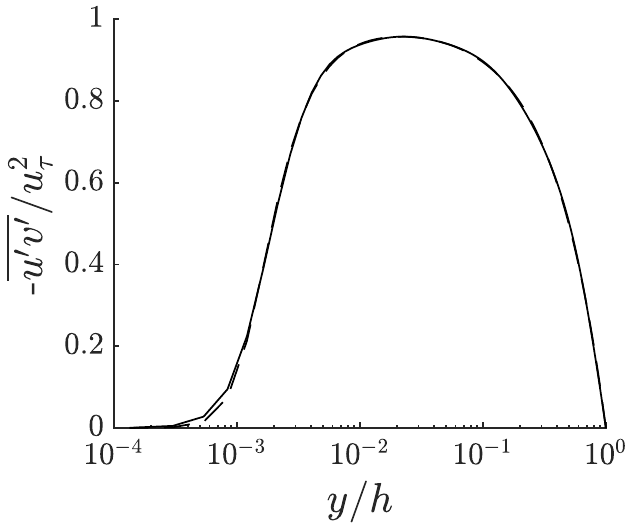}};
        \end{tikzpicture}
        \label{subfig:uvAccuracy}}
    \end{center}
    \caption{Outputs of quasi-linear approximation optimisation: $(a)$ the normalised spanwise weighting of Fourier modes; $(b)$ the wall-normal  Reynolds shear-stress profiles determined from the mean profile (dashed) and the fluctuating components (solid).}\label{fig:uvAccuracy}
\end{figure}

\subsection{Self-consistent determination of spanwise weighting}
With the self-similar weighting set $W_{r,k_x}(k_x/k_z)$ along the streamwise axis determined, an optimisation problem to determine the spanwise dependent weighting $W_{k_z}(k_z)$ is now considered. As the Reynolds shear stress generated by the fluctuating velocity field must be identical to that required for the mean profile \cite[see also][for further details]{Hwang2020}, 
the following optimisation problem for $W_{k_z}(k_z)$ is further formulated:
\begin{subequations}\label{eq:DDQLA}
\begin{equation}
    \min_{W_{k_z}} \frac{\left(\int_{0}^{2h} (\overline{u'v'}(y) - \mathbb{E}[u'v'](y))^2Q(y)\mathrm{d}y\right)^{0.5}}{\left(\int_{0}^{2h} (\overline{u'v'}(y))^2Q(y)\mathrm{d}y\right)^{0.5}} + \gamma \left(\int_{0}^{\infty}\left( \frac{\mathrm{d}^2W_{k_z}(k_z)}{\mathrm{d}\ln k_z\,^2}\right)^2R_{uv}(k_z)\mathrm{d}k_z\right)^{0.5}
\end{equation}
subject to 
\begin{equation}\label{subeq:DDQLAW}
    W_{k_z}(k_z) \geq 0,
\end{equation}
and the values of the weight at the smallest and largest considered spanwise lengths are also constrained to be zero with $W_{k_z}(\lambda_z^+ = 10) = W_{k_z}(\lambda_z = 10h) = 0$.
Here, the Reynolds shear stress $\overline{u'v'}(y)$ is given by \eqref{eq:meanUV} with the assumption that the mean velocity profile is empirically known (e.g. from (\ref{subeq:ReynoldsEddyVisc}) and (\ref{eq:meanUV})), while the Reynolds shear stress generated by the fluctuation equation (\ref{subeq:fluctuatingVelocity}) with the nonlinear term model (\ref{subeq:modelNLterm}) is denoted by $\mathbb{E}[u'v'](y)$ with the definition of
\begin{eqnarray}
    &&\mathbb{E}[u'v'](y) =\\ \nonumber
    &&\frac{1}{4\pi^2}\int_{0}^{2h}\int_{-\infty}^{\infty}\int_{-\infty}^{\infty}W_{k_z}(k_z)\sum_{r=u,v,w}W_{r,k_x}(k_x/k_z)\Phi_{uv,r}^{N_{\mathrm{POD}}}(y,y';k_x,k_z)\delta(y'-y) \mathrm{d}k_x\mathrm{d}k_z\mathrm{d}y'. 
\end{eqnarray}
The optimisation problem in (\ref{eq:DDQLA}) is weighted in logarithmic coordinates, such that equal emphasis is placed on logarithmic distance from the wall. The constraint in (\ref{subeq:DDQLAW}) ensures that the velocity covariance operators remain positive definite. 
Lastly, similar to the determination of the weighting along the streamwise wavenumber axis, a global regularisation term is considered to ensure that the weighting remains smooth, giving a physically reasonable set of velocity spectra. The smoothness regularisation is also weighted by $R_{uv}(k_z)$, where 
\begin{equation}
    R_{uv}(k_z) = \frac{1}{2\pi}\int_{h}^{2h}\int_{-\infty}^{\infty}\sum_{r=u,v,w}W_{r,k_x}(k_x/k_z)\Phi_{uv,r}^{N_{\mathrm{POD}}}(y,y';k_x,k_z)\delta(y'-y) \mathrm{d}k_x\mathrm{d}y'. 
\end{equation}
The $R_{uv}(k_z)$ tends to have large values at large spanwise wavelengths (or small $k_z$), and this is an outcome of the optimisation problem in (\ref{eq:entireOptizationProblem}). Since $\Phi_{uv}^{\mathrm{DNS}}$ contains large energy at large spanwise wavelengths, the regularisation term is more weighted at such wavelengths. This weighting accounts for the rapid decay in the Reynolds shear stress spectra observed in the previous studies \cite[see][for a further discussion]{Skouloudis2021ScalingApproximation}. It prevents erroneous behaviour in the velocity spectra at larger scales, encouraging a smoothly attached compact support at the large scales (figure \ref{subfig:SpanWeight}). 
\end{subequations}

The optimisation problem in (\ref{eq:DDQLA}) is discretised and rearranged to a standard form of a second-order cone program and solved with the MOSEK solver. Discretisation along the spanwise wavenumber axis was carried out with logarithmic spacing to maintain $\Delta(\ln(k_z))\leq 0.05$, with integration performed with the trapezoidal rule. Table \ref{tab:errorsTable} shows the number of wavenumbers, collocation points and the errors in the $Q$ and $L_{2}$ norms of problem (\ref{eq:DDQLA}). Figure \ref{fig:uvAccuracy} shows the weight determined by solving \eqref{eq:DDQLA} and the associated Reynolds shear-stress profile compared with $-\overline{u'v'}$. An almost perfect match exists between the two Reynolds shear-stress profiles, with the weighting that determines the Reynolds shear stress cospectra to provide a self-consistent approximation. With this procedure established, \S\ref{sec:DDQLAre5200} compares the results between the DQLA and a DNS \citep{Lee2015}. The predictive capabilities of the framework are also assessed by using the selected self-similar streamwise weightings as a universal weighting across a range of Reynolds numbers from $Re_\tau=10^3$ to $Re_\tau=10^5$ in \S\ref{sec:DDQLAscaling}.

\begin{table}
    \centering
    \def~{\hphantom{0}}
    \begin{tabular}{ccccccc}
    \hline
        $Re_\tau$ & $N_y$ & $N_{k_x}$  & $N_{k_z}$ & $\gamma$  & $||\overline{u'v'} - \mathbb{E}[u'v'] ||^2_{Q}$ & $||\overline{u'v'} - \mathbb{E}[u'v'] ||^2_{L_{2}}$  \\ \hline
        ~~~500 & ~128 & 172 & 126 & $1.0 \times 10^{-4}$ & $4.91 \times 10^{-4}$ & $3.03\times 10^{-5}$\\ 
        ~~1000 & ~256 & 186 & 140 & $5.0\times 10^{-5}$ & $6.17\times 10^{-4}$ & $3.12\times 10^{-5}$\\ 
        ~~2000 & ~256 & 200 & 154 & $9.0\times 10^{-5}$ & $7.30\times 10^{-4}$& $3.54\times 10^{-5}$\\ 
        ~~5200 & ~384 & 219 & 173 & $2.0\times 10^{-4}$ & $7.92\times 10^{-4}$ & $3.24\times 10^{-5}$\\ 
        ~10000 & ~512 & 232 & 186 & $1.0 \times 10^{-4}$ & $7.98\times 10^{-4}$ & $3.12\times 10^{-5}$\\ 
        ~20000 & ~768 & 246 & 200 & $7.5 \times 10^{-4}$  & $7.83 \times 10^{-4}$ & $2.97 \times 10^{-5}$  \\ 
        ~50000 & 1024 & 264 & 218 & $1.5\times 10^{-4}$& $8.53 \times 10^{-4}$  & $3.82 \times 10^{-5}$  \\ 
     100000 & 1536 & 278 & 232 & $9.0 \times 10^{-5}$ & $9.57 \times 10^{-4}$  & $2.04 \times 10^{-5}$ \\ \hline
    \end{tabular}
    \caption{Numerical and optimisation parameters used in the present study: $N_{k_x}$, the number of streamwise wavenumbers; $N_{k_z}$, the number of spanwise wavenumbers; $N_{y}$, the number of wall-normal collocation points.}
    \label{tab:errorsTable}
\end{table}

\subsection{Summary}\label{sec:IIE}
Thus far, a quasi-linear approximation has been formulated, augmented by DNS data at $Re_\tau\simeq 5200$, combined with the attached eddy hypothesis: i.e. a data-driven quasi-linear approximation. Particular efforts are given such that the model retains a `predictive' (or `extrapolative') nature for some of the key turbulence statistics and spectra at any relevant Reynolds numbers (see also \S\ref{sec:DDQLAre5200} and \ref{sec:DDQLAscaling}) with `minimal inputs': i.e. a mean velocity profile and a self-similar weight. However, in doing so, a few ad-hoc assumptions have become unavoidable for the DQLA to retain the `predictive' nature. In this respect, it would be worth documenting some of its expected characteristics originating from the construction of the model. 
\begin{enumerate}
    \item[1)] Predictability: The main feature of the DQLA is that it is predictable (or extrapolatable) for turbulence statistics and spectra at different Reynolds numbers only with two inputs: i.e. a mean profile and self-similar weight.  Note that the mean profile is empirically well documented and commonly available \cite[e.g.][]{Cess1958AFlow}, and the self-similar streamwise weight needs to be determined only once at a sufficiently high Reynolds number. By doing so, the DQLA can approach any Reynolds numbers without additional DNS data as long as the flow remains fully turbulent. Importantly, as is seen in \S \ref{sec:DDQLAscaling}, most of the known Reynolds-number-dependent scaling behaviours of turbulence statistics and spectra appear to be reproduced by the present DQLA. Therefore, it may be a valuable tool for the studying turbulence statistics at extremely high Reynolds numbers, where an accurate data set is challenging to obtain, as demonstrated by \cite{Skouloudis2021ScalingApproximation} and \S\ref{sec:DDQLAscaling}.
    \item[2)] Performance: In the DQLA, the first input, a mean profile, is assumed to be known at a given Reynolds number and is determined by an eddy viscosity closure, giving the law of the wall. Given the robustness of the law of the wall, this input should result in reasonable approximations at extremely high Reynolds numbers. The second input used here is the streamwise weighting. The DQLA leverages the self-similar nature of the eddies in the attached eddy hypothesis, focusing on modelling the logarithmic layer. Given the significance of the logarithmic layer grows with $Re_\tau$, the modelling framework should perform best for the high Reynolds numbers, where the features not modelled by this self-similar weight bear a minor significance: e.g. near-wall motions.
    \item[3)] Limitations: The extrapolation capability of the DQLA is, however, obtained at the cost of the accuracy of the resulting turbulence statistics, especially compared to the recent data-driven modelling efforts \cite[]{Zare2017ColourTurbulence,abootorabi_zare_2023,Holford2022}. In particular, the physical processes and the related velocity spectra originating from the original nonlinear term in (\ref{subeq:exactNLterm}) (i.e. streak instability/transient-growth and energy cascade) were ignored entirely. However, it should be pointed out that this is a fundamental limitation of any modelling efforts of turbulent fluctuations based on the current linearised Navier-Stokes equations. Even if highly accurate turbulence statistics are obtained with a more accurate forcing incorporating the ignored part of the spectra, the model nonlinear term in (\ref{subeq:modelNLterm}) or any of its variants can only be phenomenological, and they do not have the dynamics of the original nonlinear term in (\ref{subeq:exactNLterm}). As such, it is important to remember that the DQLA is a model primarily for turbulence statistics, leveraging the physical processes that the linearised Navier-Stokes equations can be depict. In this respect, the models that describe better turbulent `dynamics' may be found from some of the recent studies employing more sophisticated state decompositions of the velocity field \cite[e.g.][]{farrell_ioannou_2012,ThomasEtAl2015,farrell_ioannou_jiménez_constantinou_lozano-durán_nikolaidis_2016,hernández_yang_hwang_2022,hernández_yang_hwang_2022_2}.
\end{enumerate}

\section{Data-driven quasi-linear approximation at $Re_\tau = 5200$}\label{sec:DDQLAre5200}
\begin{figure}[h]
    \begin{center}
    \sidesubfloat[]{\hspace{-0.5cm}
        \begin{tikzpicture}
            \node (img)  {\includegraphics[width = 5.0cm]{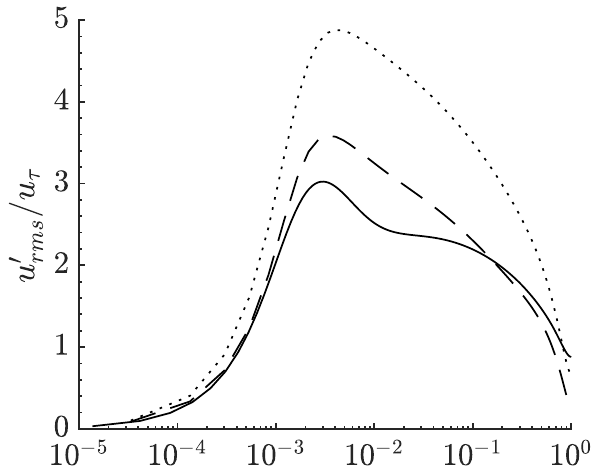}};
        \end{tikzpicture}
        \label{subfig:CompareUU}}
    \sidesubfloat[]{\hspace{-0.5cm}
        \begin{tikzpicture}
          \node (img)  {\includegraphics[width = 5.2cm]{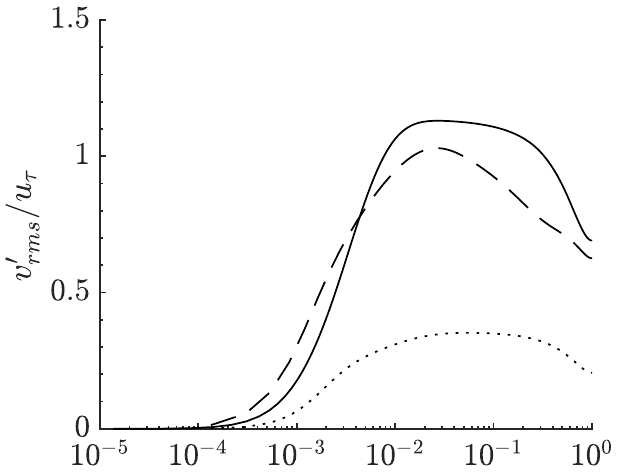}};
        \end{tikzpicture}
        \label{subfig:CompareVV}} \hfill \\
    \sidesubfloat[]{\hspace{-0.5cm}
        \begin{tikzpicture}
            \node (img)  {\includegraphics[width = 5.0cm]{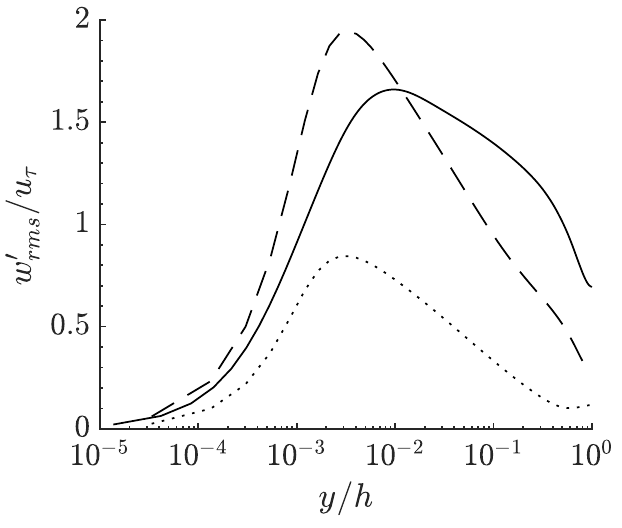}};
        \end{tikzpicture}
        \label{subfig:CompareWW}}
    \sidesubfloat[]{\hspace{-0.5cm}
        \begin{tikzpicture}
          \node (img)  {\includegraphics[width = 5.1cm]{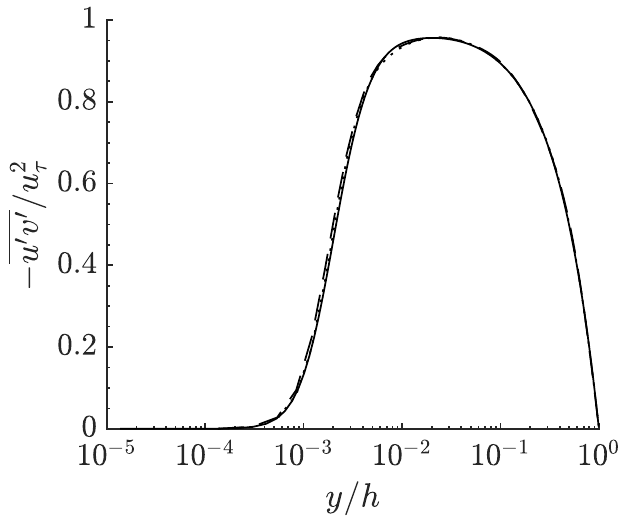}};
        \end{tikzpicture}
        \label{subfig:CompareUV}}
    \end{center}
    \caption{Comparison between DNS at $Re_\tau = 5186$ (solid), DQLA (dashed) and MQLA (dotted) at $Re_\tau = 5200$. $(a)$ streamwise, $(b)$ wall-normal and $(c)$ spanwise root mean square velocity and $(d)$ Reynolds shear stress profiles.}\label{fig:RMScomparison}
\end{figure}

\subsection{One-point turbulence statistics}
The DQLA is first performed at $Re_\tau = 5200$ for comparison purposes with DNS data, as well as to compare and evaluate the effects of including $k_x \neq 0$ Fourier modes in contrast to the MQLA considering only $k_x=0$ Fourier modes. Here, the MQLA has been reevaluated using the optimisation problem \eqref{eq:DDQLA}, with $\mathbb{E}[u'v']$ determined for streamwise uniform modes ($k_x = 0$). 
Figure \ref{fig:RMScomparison} plots the root mean square (rms) velocity fluctuation and Reynolds shear stress profiles for the DNS, DQLA and MQLA. Reynolds shear stress profiles are virtually identical in logarithmic coordinates (fig \ref{subfig:CompareUV}; see also table \ref{tab:errorsTable}), indicating successful implementation of the optimisation problem (\ref{eq:DDQLA}) for DQLA and MQLA. 
The rms velocity profiles of the DQLA show significant quantitative improvement over the MQLA (figures \ref{fig:RMScomparison}a-c). The overall effects of including $k_x \neq 0$ Fourier modes reduces the anisotropy in the profiles from the DQLA to be more consistent with the DNS profiles. For instance, $({u'_{\mathrm{rms,max}}}/{w'_{\mathrm{rms,max}}},{u'_{\mathrm{rms,max}}}/{v'_{\mathrm{rms,max}}})$ is approximately (1.82, 2.67), (1.84, 3.48) and (5.78, 13.9) in the DNS, DQLA and MQLA, respectively. There is a strong agreement between the DNS and DQLA ratios for the wall-parallel velocity components, with the inclusion of $k_x \neq 0$ modes alleviating this anisotropy. 

This mismatch in the MQLA is evidently due to considering only $k_x=0$ Fourier modes. For the $k_x = 0$ case, the wall-normal derivative of wall-normal velocity fluctuations must exclusively be balanced by the spanwise variation in the spanwise velocity spectra, given the form of the continuity equation: i.e.
\begin{equation}
    \frac{\mathrm{d}\hat{v}'}{\mathrm{d}y}+ik_z \hat{w}'=0.
\end{equation}
Hence, the spanwise velocity can be determined directly from the wall-normal velocity, itself determined solely from the Orr-Sommerfeld equation -- note that the Orr-Sommerfeld equation for the wall-normal velocity is not coupled with the streamwise and spanwise velocity. Similarly, when $k_x=0$, the streamwise momentum equation is only passively coupled with the other two momentum equations through the wall-normal velocity: i.e.
\begin{equation}
    \frac{\partial \hat{u}'}{\partial t}+\hat{v}'\frac{\mathrm{d}U}{\mathrm{d}y}=\frac{\mathrm{d}\nu_T}{\mathrm{d}y}\hat{u}' + \nu_T\Delta_{y,z}\hat{u} + \hat{f}_u'
\end{equation}
where $\Delta_{y,z} = \frac{\partial^2}{\partial y^2} -k_z^2$. However, this is the only momentum equation containing the mean shear, the source of the energy production in the linearised fluctuation equations. Therefore, when only $k_x=0$ Fourier modes are considered for velocity fluctuations, like in the MQLA, there is no way to transport the fluctuation energy produced by the mean shear at the streamwise velocity component to the other two components. This results in the highly overpredicted streamwise-to-spanwise/wall-normal rms velocity ratios in the MQLA. 
By relaxing the model to accommodate $k_x \neq 0$ modes in the DQLA, the streamwise momentum equation is now fully coupled with the equations for the other two components through continuity and pressure, significantly reducing the odd anisotropy in the rms velocity profiles observed in the MQLA. While the ratio of the peaks of the wall-parallel components is quantitatively similar between the DNS and the DQLA, the ratios in the streamwise and wall-normal components are still overpredicted in the DQLA. This is presumably due to the crude model for the nonlinear term in (\ref{subeq:modelNLterm}) and the consideration of only two leading POD modes for the construction of velocity spectra, which are not able to incorporate some of the key nonlinear processes of the flow, such as the streak instability or transient growth in the self-sustaining process \citep{Hamilton1995,Schoppa2002CoherentTurbulence,deGiovanetti2017StreakMotions,Lozano-Duran2021Cause-and-effectTurbulence} and the resulting energy cascade \citep{Doohan2021MinimalTurbulence}.  


\begin{figure}
    \begin{center}
    \sidesubfloat[]{\hspace{-0.5cm}
        \begin{tikzpicture}
            \node (img)  {\includegraphics[width = 4.30cm]{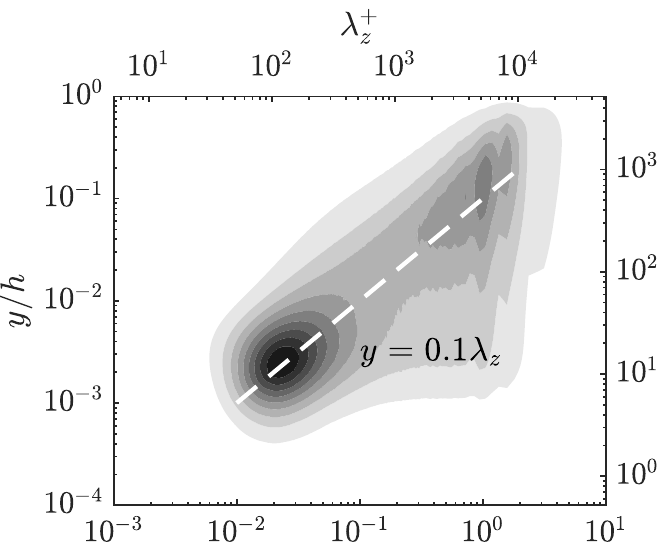}};
        \end{tikzpicture}
        \label{subfig:kzPhiuuDNS}}
    \sidesubfloat[]{\hspace{-0.5cm}
        \begin{tikzpicture}
            \node (img)  {\includegraphics[width = 4.10cm]{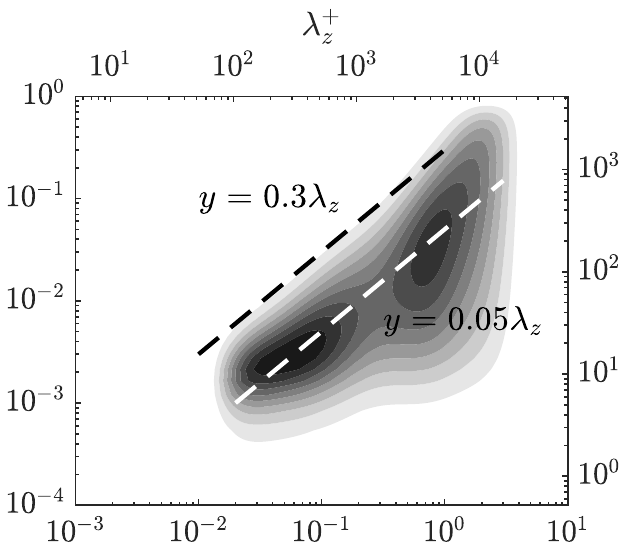}};
        \end{tikzpicture}
        \label{subfig:kzPhiuuQLA}} 
    \sidesubfloat[]{\hspace{-0.5cm}
        \begin{tikzpicture}
            \node (img)  {\includegraphics[width = 4.40cm]{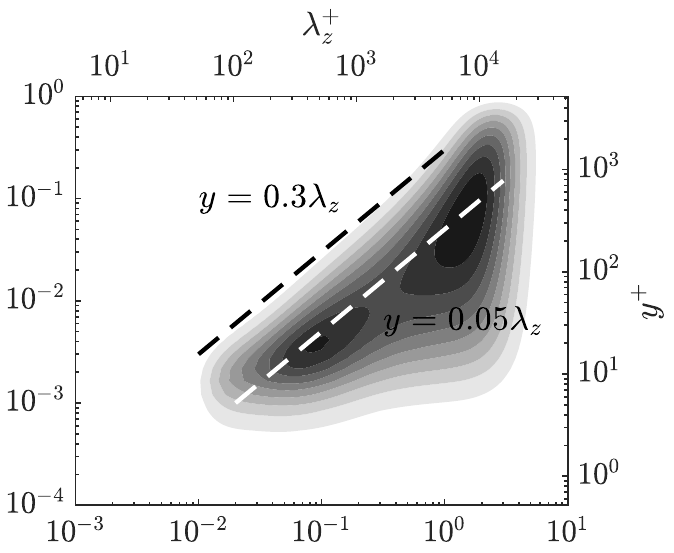}};
        \end{tikzpicture}
        \label{subfig:kzPhiuuMQLA}}   \hfill \\
        \sidesubfloat[]{\hspace{-0.5cm}
        \begin{tikzpicture}
            \node (img)  {\includegraphics[width = 4.30cm]{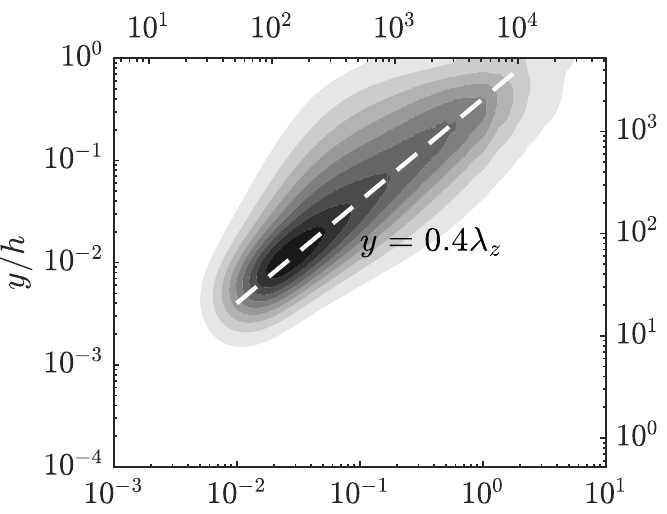}};
        \end{tikzpicture}
        \label{subfig:kzPhivvDNS}}
    \sidesubfloat[]{\hspace{-0.5cm}
        \begin{tikzpicture}
            \node (img)  {\includegraphics[width = 4.10cm]{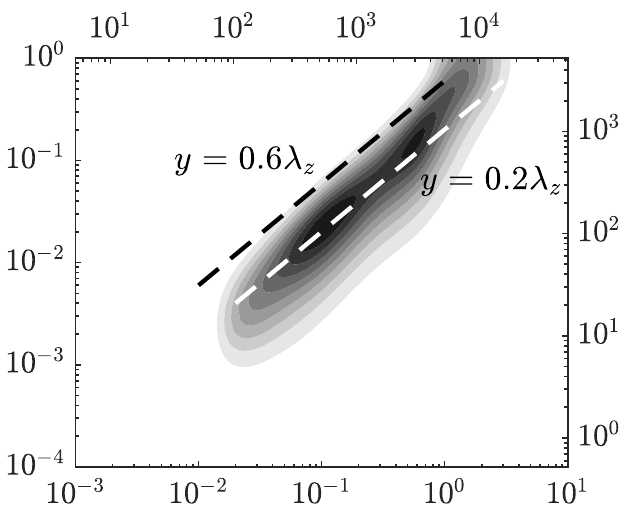}};
        \end{tikzpicture}
        \label{subfig:kzPhivvQLA}} 
    \sidesubfloat[]{\hspace{-0.5cm}
        \begin{tikzpicture}
            \node (img)  {\includegraphics[width = 4.40cm]{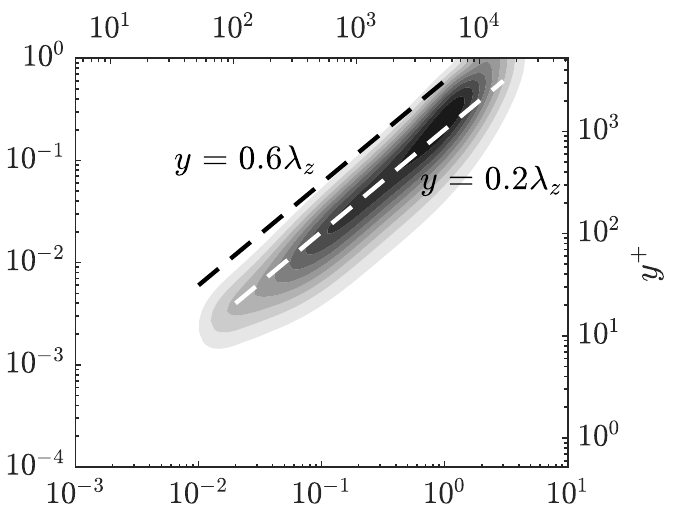}};
        \end{tikzpicture}
        \label{subfig:kzPhivvMQLA}}   \hfill \\
            \sidesubfloat[]{\hspace{-0.5cm}
        \begin{tikzpicture}
            \node (img)  {\includegraphics[width = 4.30cm]{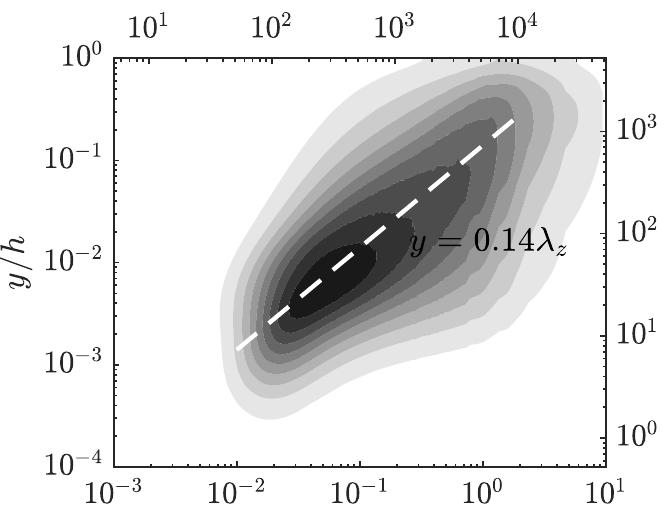}};
        \end{tikzpicture}
        \label{subfig:kzPhiwwDNS}}
    \sidesubfloat[]{\hspace{-0.5cm}
        \begin{tikzpicture}
            \node (img)  {\includegraphics[width = 4.10cm]{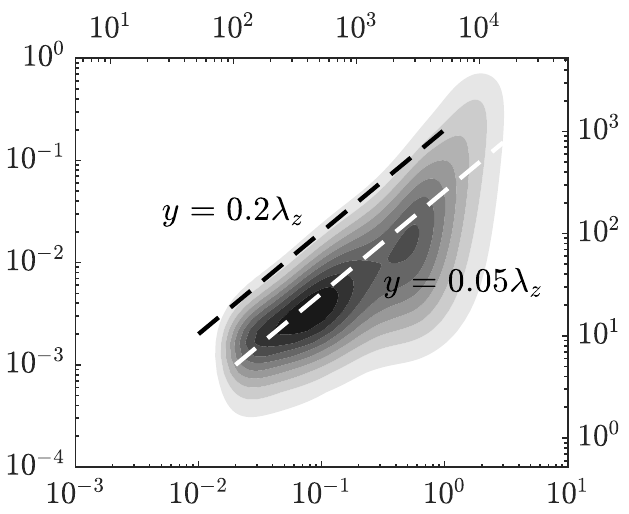}};
        \end{tikzpicture}
        \label{subfig:kzPhiwwQLA}} 
    \sidesubfloat[]{\hspace{-0.5cm}
        \begin{tikzpicture}
            \node (img)  {\includegraphics[width = 4.40cm]{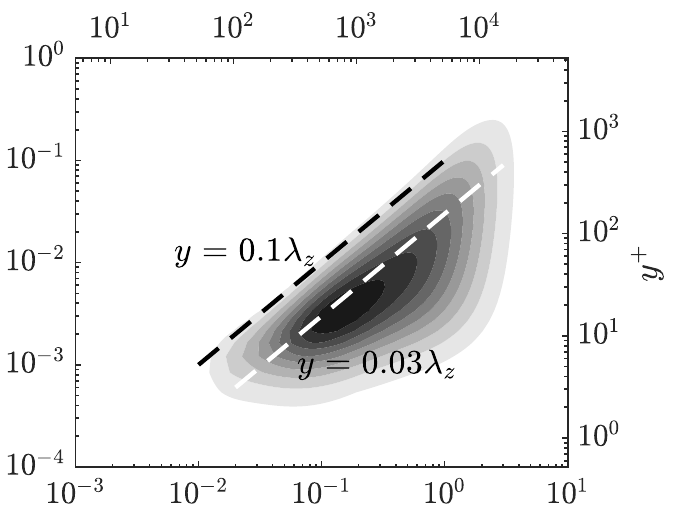}};
        \end{tikzpicture}
        \label{subfig:kzPhiwwMQLA}}   \hfill \\
            \sidesubfloat[]{\hspace{-0.5cm}
        \begin{tikzpicture}
            \node (img)  {\includegraphics[width = 4.30cm]{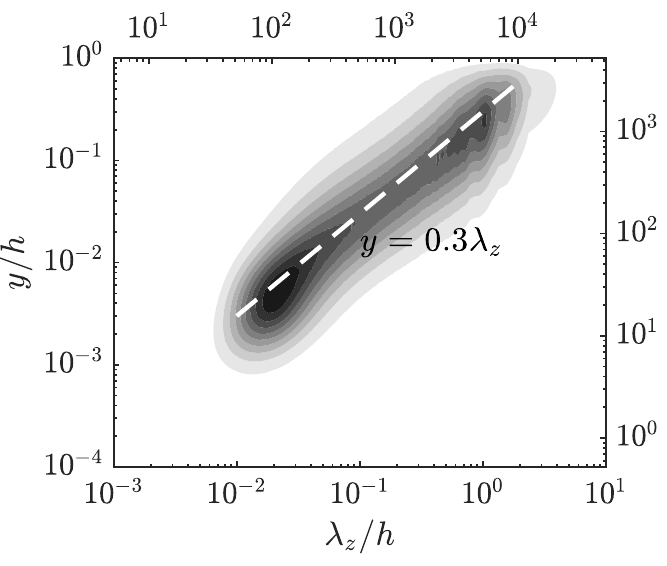}};
        \end{tikzpicture}
        \label{subfig:kzPhiuvDNS}}
    \sidesubfloat[]{\hspace{-0.5cm}
        \begin{tikzpicture}
            \node (img)  {\includegraphics[width = 4.10cm]{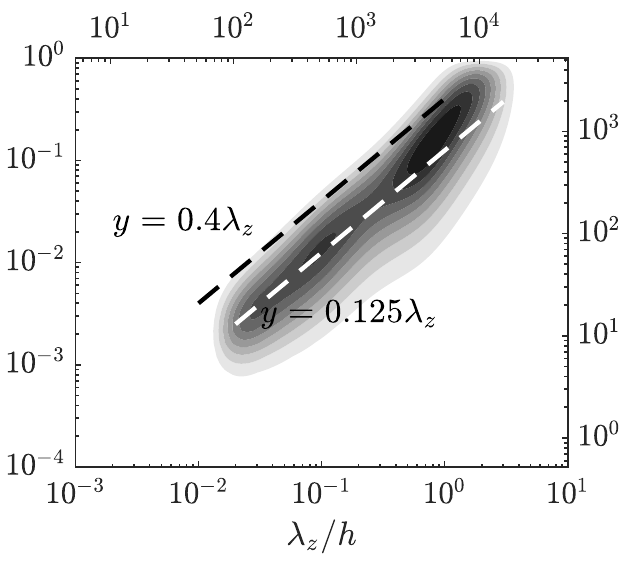}};
        \end{tikzpicture}
        \label{subfig:kzPhiuvQLA}} 
    \sidesubfloat[]{\hspace{-0.5cm}
        \begin{tikzpicture}
            \node (img)  {\includegraphics[width = 4.40cm]{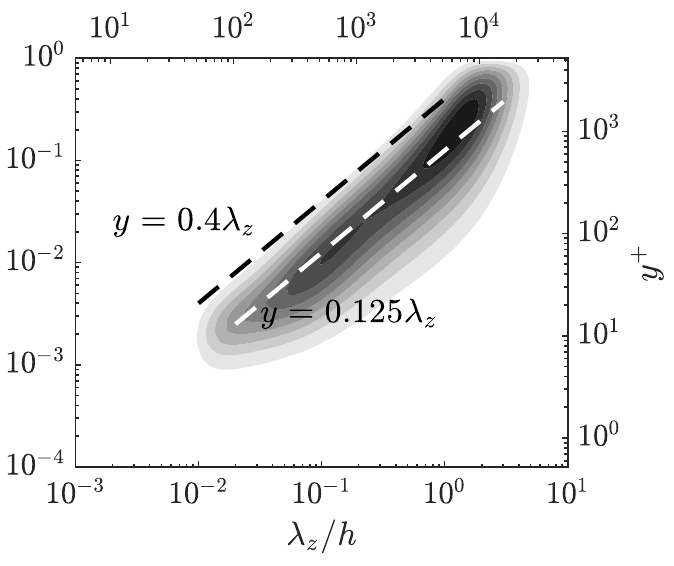}};
        \end{tikzpicture}
        \label{subfig:kzPhiuvMQLA}} \hfill
    \end{center}
    \caption{Premultiplied spanwise wavenumber spectra from (a,d,g,j) DNS at $Re_\tau = 5186$, (b,e,h,k) the DQLA  and (c,f,i,l) the MQLA at $Re_\tau = 5200$: (a,b,c) streamwise velocity; (d,e,f) wall-normal velocity; (g,h,i) spanwise velocity; (j,k,l) Reynold shear stress. The contours are normalised by 0.1 times the maximum value.}\label{fig:spanwiseSpectraRe5200}
\end{figure}

\subsection{One-dimensional spectra}
Figure \ref{fig:spanwiseSpectraRe5200} compares the spanwise one-dimensional velocity spectra from DNS with those of the DQLA and MQLA. Figures \ref{fig:spanwiseSpectraRe5200}(a-c) show the streamwise velocity spectra for the DNS, DQLA and MQLA, respectively. The DQLA and MQLA are energetic along the same linear ridge (the white-dashed line $y = 0.05\lambda_z$), with the energetic content dropping off approximately along the same linear length scale (the black-dashed line $y= 0.3\lambda_z$). 
There is a good qualitative agreement between all three of the spectra in the modelling of the attached features of the $y$- and $h$-scaling motions for $\lambda_z/h \gtrsim 0.1$: both the DQLA and MQLA streamwise velocity spectra for $\lambda_z/h \gtrsim 0.1$ penetrate the near-wall region below $y^+\lesssim 100$, where the wall-normal velocity spectra are not energetic due to the boundary condition (see figures \ref{fig:spanwiseSpectraRe5200}d-f). The main qualitative difference is the location of the near-wall peak. In the DQLA, this near-wall peak occurs for $150 \lesssim  \lambda_z^+ \lesssim 400$, whereas in the MQLA, the near-wall peak occurs at larger spanwise length scales $300 \lesssim  \lambda_z^+ \lesssim 500$. These near-wall peaks occur at larger spanwise length scales than the DNS, which occurs for $100 \lesssim  \lambda_z^+ \lesssim 150$, with all of them coinciding at $y^+ \approx 10$. Another common difference when comparing the DQLA and MQLA to DNS is the relative strength of the outer peak. In both the MQLA and DQLA, the relative strength of the outer peak is significantly greater than the DNS counterpart. However, the location of these outer peaks is consistent with the DNS relative to its energetic linear ridge, with the outer peak occurring at $\lambda_z/h \approx 1$ on $y = 0.05 \lambda_z$ in the DQLA and MQLA, and $y = 0.1\lambda_z$ in the DNS, with the DQLA and MQLA being energetic slightly closer to the wall. 

Figures \ref{fig:spanwiseSpectraRe5200}(d-f) compare the wall-normal velocity spectra. There is little qualitative difference between the DQLA and MQLA again, with both  sharing qualitative features of the DNS spectra. Like the streamwise velocity spectra, the DQLA and MQLA are more energetic closer to the wall along $y = 0.2\lambda_z$ as opposed to $y = 0.4\lambda_z$ seen in DNS. The location of the peak in the DQLA, relative to the MQLA peak, also has the same trend as the near-wall peak location in the streamwise velocity spectra. The DQLA wall-normal velocity spectra are more energetic lower down on the energetic linear ridge ($\lambda_z/h \approx 0.1$ on $y = 0.2\lambda_z$). While this shifts the peak in the wall-normal velocity spectra towards that of the DNS location ($\lambda_z/h \approx 0.02$), the DQLA peak behaviour still occurs at larger length scales than that observed in the DNS. 

Figures \ref{fig:spanwiseSpectraRe5200}(g-i) compares the spanwise velocity spectra. The comparison between the MQLA and DNS is identical to that in the streamwise velocity spectra: the MQLA is energetic closer to the wall, and the peak location occurs at larger length scales. Relaxing the $k_x = 0$ consideration considerably improves the modelling when comparing the DQLA to the MQLA. 
In general, the DQLA is energetic further away from the wall, and, at the larger length scales ($\lambda_z/h \approx 1$), the DQLA spectra extend much closer to the channel centre line, more in line with the DNS spectra. 

The Reynolds shear stress spectra are compared in figures \ref{fig:spanwiseSpectraRe5200}(j-l). In this case, there is little to no difference between the MQLA and DQLA, and both compare identically to the DNS spectra. Given that the Reynolds shear stress does not develop strong spectra associated with energy cascade than the other velocity components in the inertial subrange \citep{Lumley1967SimilaritySpectrum}, the spectra of DQLA and MQLA, in which such spectra are ignored by construction, are qualitatively very similar to those of the DNS. The main qualitative difference is the location of the primary peak, with this occurring at an outer length scale in the DQLA and MQLA, compared to the primary peak of the DNS appearing in the near-wall region.

To summarise these results with attention to comparing the DQLA to the DNS data, in general, the DQLA is much more energetic closer to the wall: i.e. the wall-normal locations of the linear ridges in figure \ref{fig:spanwiseSpectraRe5200} are closer to the wall. While in \cite{Hwang2020}, this was attributed to the lack of  $k_x \neq 0$ considerations, the same differences are found in the DQLA also. These common differences appear because the streak instability or transient growth mechanism is absent in the DQLA and MQLA by construction. The length scale at which this occurs in the actual turbulent flow is further away from the wall: $\lambda_z/y\simeq 0.5-0.7$ \cite[]{Hwang2015,deGiovanetti2017StreakMotions}. This justifies the large discrepancy in the length scale of the energetic linear ridge in the wall-normal and spanwise velocity spectra, considering these velocity components should be more energetic along the length scale selection of the streak instability. In general, the DQLA appears to shift the high energetic contents of the velocity spectra towards smaller spanwise length scales, although the DQLA still does not achieve near-wall peaks at the spanwise length scales present in the DNS. This could be because the DQLA was performed using a universal self-similar weight selected in the logarithmic region. Instead, a near-wall correction might be required to improve this feature. 

\begin{figure}
    \begin{center}
    \sidesubfloat[]{\hspace{-0.5cm}
        \begin{tikzpicture}
            \node (img)  {\includegraphics[width = 4.8cm]{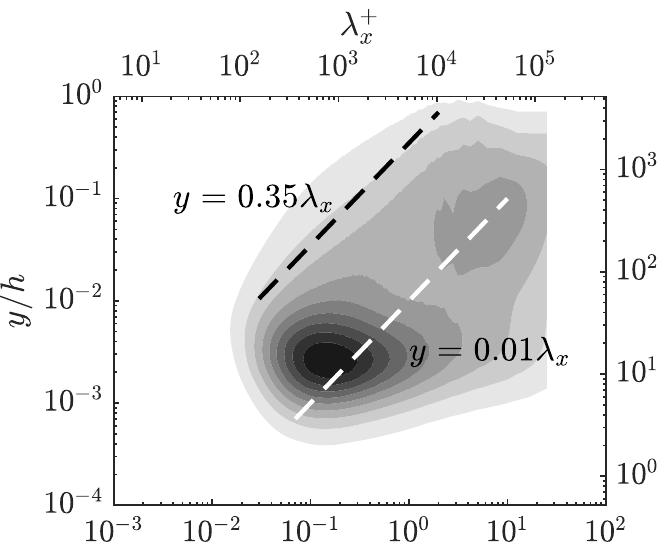}};
        \end{tikzpicture}
        \label{subfig:kxPhiuuDNS}} \hspace{1cm}
    \sidesubfloat[]{\hspace{-0.5cm}
        \begin{tikzpicture}
            \node (img)  {\includegraphics[width = 4.9cm]{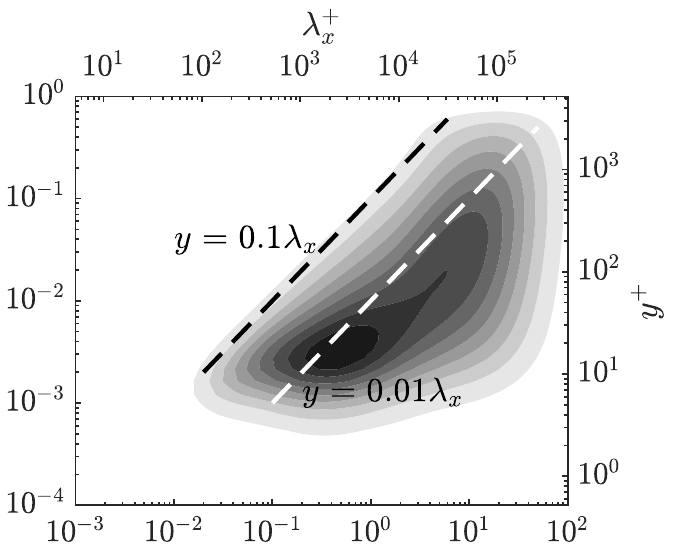}};
        \end{tikzpicture}
        \label{subfig:kxPhiuuQLA}} \hfill \\
    \sidesubfloat[]{\hspace{-0.5cm}
        \begin{tikzpicture}
            \node (img)  {\includegraphics[width = 4.8cm]{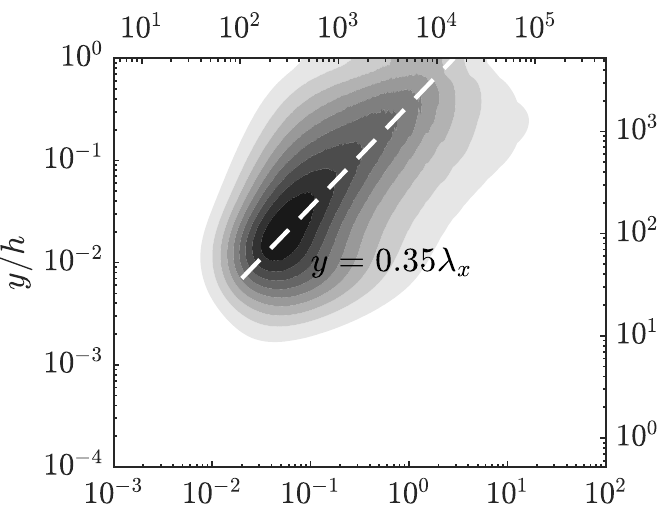}};
        \end{tikzpicture}
        \label{subfig:kxPhivvDNS}} \hspace{1cm}
    \sidesubfloat[]{\hspace{-0.5cm}
        \begin{tikzpicture}
            \node (img)  {\includegraphics[width = 4.9cm]{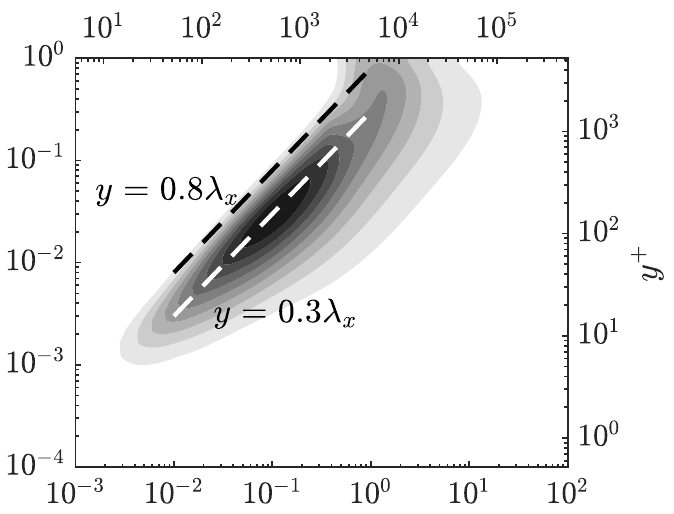}};
        \end{tikzpicture}
        \label{subfig:kxPhivvQLA}} \hfill \\
    \sidesubfloat[]{\hspace{-0.5cm}
        \begin{tikzpicture}
            \node (img)  {\includegraphics[width = 4.8cm]{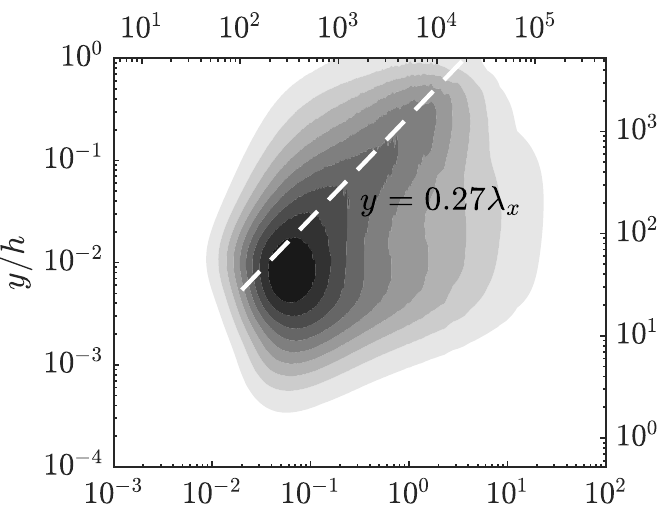}};
        \end{tikzpicture}
        \label{subfig:kxPhiwwDNS}} \hspace{1cm}
    \sidesubfloat[]{\hspace{-0.5cm}
        \begin{tikzpicture}
            \node (img)  {\includegraphics[width = 4.9cm]{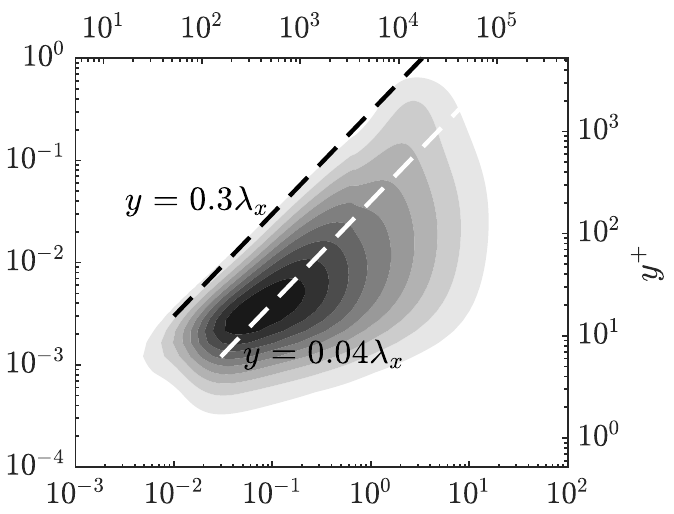}};
        \end{tikzpicture}
        \label{subfig:kxPhiwwQLA}} \hfill \\
    \sidesubfloat[]{\hspace{-0.5cm}
        \begin{tikzpicture}
            \node (img)  {\includegraphics[width = 4.8cm]{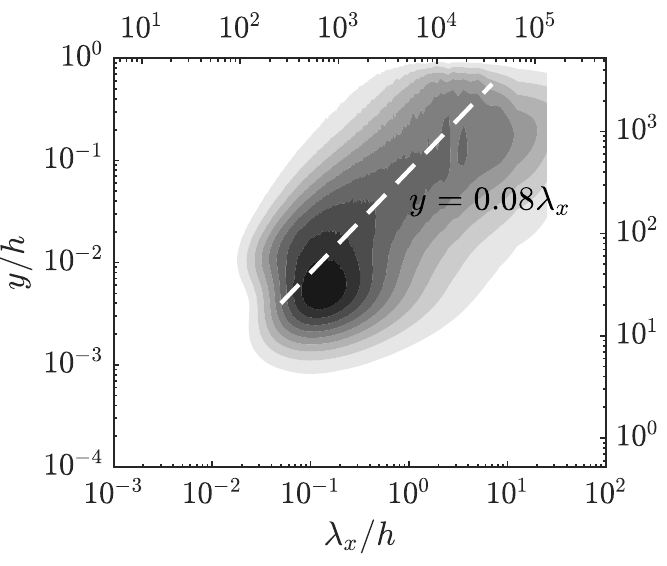}};
        \end{tikzpicture}
        \label{subfig:kxPhiuvDNS}} \hspace{1cm}
    \sidesubfloat[]{\hspace{-0.5cm}
        \begin{tikzpicture}
            \node (img)  {\includegraphics[width = 4.9cm]{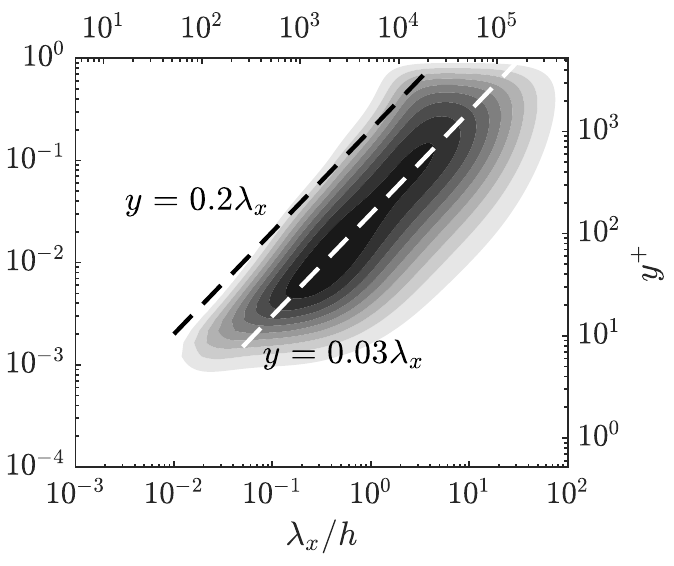}};
        \end{tikzpicture}
        \label{subfig:kxPhiuvQLA}}
    \end{center}
    \caption{Premultiplied spanwise wavenumber spectra from (a,c,e,g) DNS at $Re_\tau = 5186$ and $(b,d,f,h)$ the DQLA at $Re_\tau = 5200$: (a,b) streamwise velocity spectra; (c,d) wall-normal velocity spectra; (e,f) spanwise velocity spectra; (g,h) Reynolds shear stress spectra. The contours are normalised by 0.1 times the maximum value.}\label{fig:streamwiseSpectraKzhRe5200}
\end{figure}

Figure \ref{fig:streamwiseSpectraKzhRe5200} now compares the streamwise one-dimensional velocity and Reynolds shear stress spectra between DNS and the DQLA. Here the MQLA is not included, given $k_x \neq 0$ information is not included in the MQLA. Figures 
\ref{fig:streamwiseSpectraKzhRe5200}(a,b) compares the streamwise velocity spectra. There is a good qualitative agreement between the spectra for the long streaky features in the DNS associated with the lift-up effect, i.e. along the $y=0.01\lambda_z$ linear ridge \citep{Hwang2015}. Although the spectra from the DNS suffer from the finite streamwise domain considered, the DQLA well replicates the footprints for the large-scale structures, remaining attached over the entire range of streamwise length scales considered. Similar to the spanwise one-dimensional spectra, the location of the near-wall primary peak in the DQLA occurs at slightly larger length scales than that in the DNS ($\lambda_x^+ \approx 2000$ compared to $\lambda_x^+ \approx 1000$). The most noticeable qualitative difference in the streamwise velocity spectra occurs above the $y\approx0.1 \lambda_x$ linear ridge. In the DQLA, this linear ridge describes a cutoff point for the energy-containing motions, above which the energetic contents drop to zero. On the other hand, in the DNS, another energetic linear ridge (i.e. $y=0.35 \lambda_x$ in figure \ref{fig:streamwiseSpectraKzhRe5200}a), strongly correlated with wall-normal and spanwise velocity field (see also figures \ref{fig:streamwiseSpectraKzhRe5200}c,e), is present, as discussed in detail in \cite{Hwang2015}. This ridge is associated with vortex packets emerging from the instability or transient growth of long streaky motions \citep{deGiovanetti2017StreakMotions}, the feature entirely neglected in the DQLA by the construction detailed in \S\ref{sec:IIC}.


Figures \ref{fig:streamwiseSpectraKzhRe5200}(c,d) compares the wall-normal velocity spectra. There is a good qualitative agreement between the spectra with respect to the energy-containing motions, with a single linear ridge occurring in both at the same approximate length scale, $y\sim 0.35\lambda_x$ and $y\sim 0.30\lambda_x$ in the DNS and DQLA, respectively. Overall, the streak-instability length scale selection mentioned above is phenomenologically well mimicked by the self-similar weighting in the DQLA, providing a reasonable approximation for a majority of the velocity spectra. 
The main qualitative difference is that the DNS velocity spectra are more energetic above the linear ridge than the DQLA, which has a linear cutoff along $y\approx 0.8\lambda_z$. Also, in the DNS spectra, the contour lines with low levels for  $y> 0.8\lambda_x$ do not follow any linear scaling like $y \sim \lambda_x$, and this part has previously been associated with energy cascade developing a $k_x^{-5/3}$ spectrum \cite[e.g.][]{Agostini2017}. This is expected, given the determination of the self-similar streamwise weighting effectively removed features associated with the energy cascade. Similar to the spanwise one-dimensional velocity spectra, the primary peak in the DQLA occurs at larger length scales than the DNS. The wall-normal velocity spectra in the DQLA also fall off at a slower rate below this primary peak when compared to the DNS. The low-level contours in the DQLA follow the linear ridge, extending to $y^+ \approx 1$ for $\lambda_x^+ \approx 2$, whereas the DNS extends to  $y^+ \approx 1$ for $\lambda_x^+ \approx 100$. This discrepancy in peak location and the DQLA extending to the near-wall region for small streamwise length scales is presumably from the use of a self-similar weighting across all length scales, and is likely contributing to the overprediction in the near-wall wall-normal velocity intensity profile and Reynolds shear stress profile. A near-wall correction to the modelling procedure would be required to alleviate this. Although the overall effects on the wall-normal turbulence intensity and Reynolds shear stress profiles are expected to be minor, given that most of the energy is contained at log-layer associated length scales at high Reynolds numbers.

The spanwise velocity spectra are compared in figures \ref{fig:streamwiseSpectraKzhRe5200}(e,f), with the DQLA spectra sharing much of the qualitative features of the streamwise velocity spectra from the DQLA (figure \ref{fig:streamwiseSpectraKzhRe5200}b). The DQLA models the attached footprint and provides the approximate location of the near-wall peak in the streamwise direction, albeit at a slightly larger length scale again ($\lambda_x^+ \sim 500$ in the DQLA compared to $\lambda_x^+ \sim 300$). Most noticeably of all the velocity spectra, the main energetic regions in spectra are much closer to the wall than the DNS. The DNS velocity spectra show that the main energy-containing regions are associated with the streak instability along $y \sim 0.27 \lambda_x$. In the DQLA, this appears to be absent, given that the streak instability (and/or transient growth) mechanism has been associated with the spanwise velocity fluctuation \citep{Schoppa2002CoherentTurbulence,deGiovanetti2017StreakMotions} leading to the streamwise meandering motion of streaks \citep{Hutchins2007EvidenceLayers}. 


The Reynolds shear stress spectra is compared in figures \ref{fig:streamwiseSpectraKzhRe5200}(g,h), with the main observations consistent with the comparisons of the streamwise and wall-normal velocity spectra. The Reynolds shear stress spectra generated by the DQLA is energetic closer to the wall when compared to the DNS, with most of the energetic content following $y \sim 0.03\lambda_x$ in the DQLA versus $y \sim 0.08\lambda_x$ in the DNS. Similar to the wall-normal velocity spectra, the Reynolds shear stress spectra extends along the linear ridge to smaller length scales when compared to DNS, leading to the slight over prediction in the Reynolds stress profile (figure \ref{fig:RMScomparison}d). Given the nature of the one-way coupling in the current linear model, with the wall-normal velocity spectra determined independently of the other components, the differences in Reynolds shear stress profile again highlight the lack of the streak instability or transient growth mechanism in the DQLA. In the DQLA model, a majority of the Reynolds shear stress is generated through the lift-up effect, with the wall-normal velocity spectra driving the generation of the streamwise velocity spectra, with the Reynolds shear stress generated closer to the wall. In reality, wall-normal and streamwise velocity fluctuations can be correlated further away from the wall, through an instability or transient growth in the streamwise fluctuations (i.e. the streak instability and transient growth) feeding into the wall-normal ones. Given the current model uses POD modes to construct the velocity spectra and implicitly vary the forcing in the wall-normal direction, a more carefully prescribed colour is required to mimic the streak instability and the subsequent Reynolds shear stresses generated away from the wall.

\begin{figure}
    \begin{center}
    \sidesubfloat[]{\hspace{-0.5cm}
        \begin{tikzpicture}
            \node (img)  {\includegraphics[width = 5.0cm]{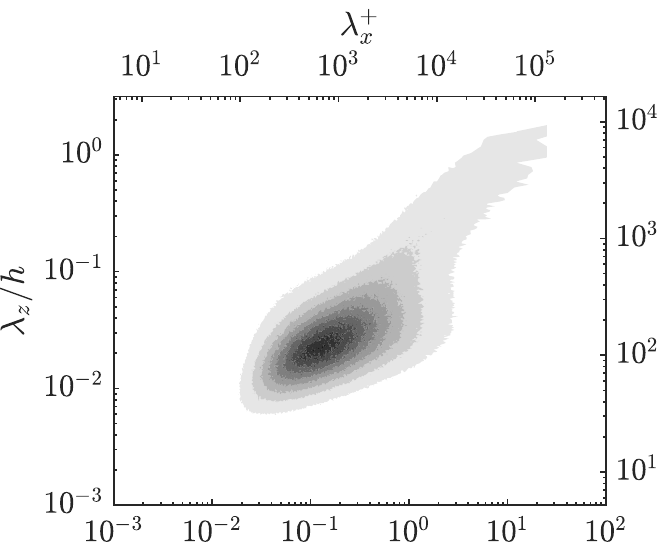}};
        \end{tikzpicture}
        \label{subfig:kxkzPhiuuDNS}}
    \sidesubfloat[]{
        \begin{tikzpicture}
            \node (img)  {\includegraphics[width = 5.1cm]{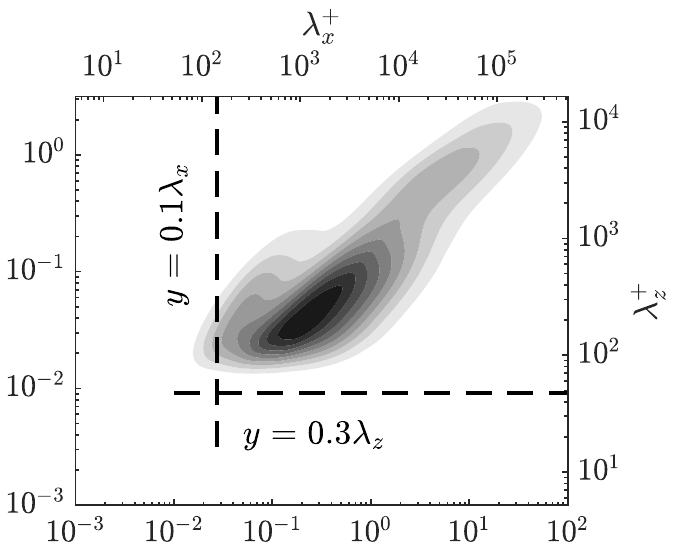}};
        \end{tikzpicture}
        \label{subfig:kxkzPhiuuQLA}} \hfill \\
    \sidesubfloat[]{\hspace{-0.5cm}
        \begin{tikzpicture}
            \node (img)  {\includegraphics[width = 5.0cm]{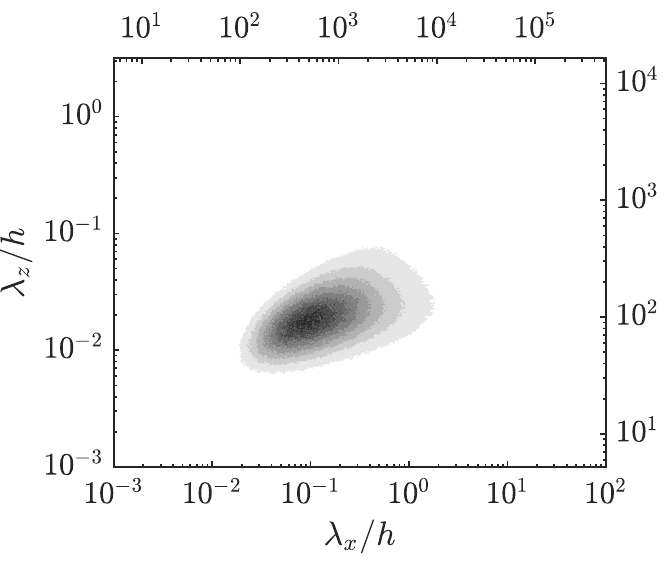}};
        \end{tikzpicture}
        \label{subfig:kxkzPhiuvDNS}}
    \sidesubfloat[]{
        \begin{tikzpicture}
            \node (img)  {\includegraphics[width = 5.1cm]{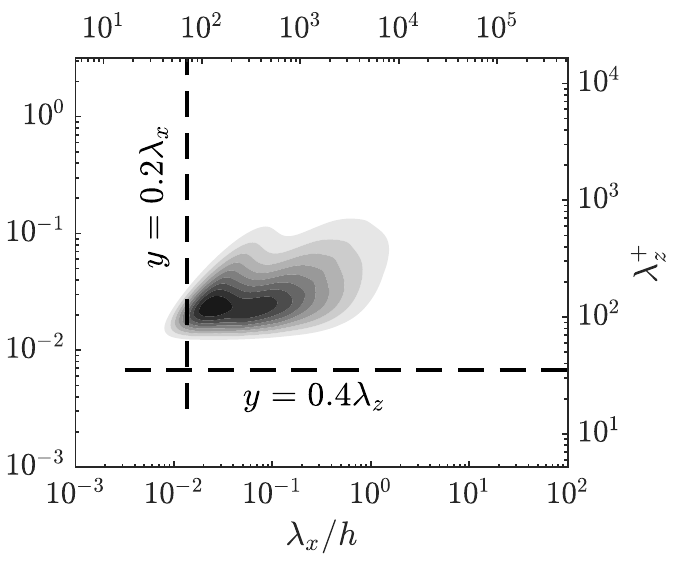}};
        \end{tikzpicture}
        \label{subfig:kxkzPhiuvQLA}}
    \end{center}
    \caption{Premultiplied two-dimensional wavenumber spectra  at $y^+ \approx 15$ from (a,c) DNS at $Re_\tau = 5186$ and (b,d) the DQLA at $Re_\tau = 5200$: (a,b) streamwise velocity spectra; (c,d) Reynolds shear stress cospectra. The contours are normalised by 0.1 times the maximum value.}\label{fig:kxkzPhiRe5200}
\end{figure}

\begin{figure}
    \begin{center}
    \sidesubfloat[]{\hspace{-0.5cm}
        \begin{tikzpicture}
            \node (img)  {\includegraphics[width = 5.3cm]{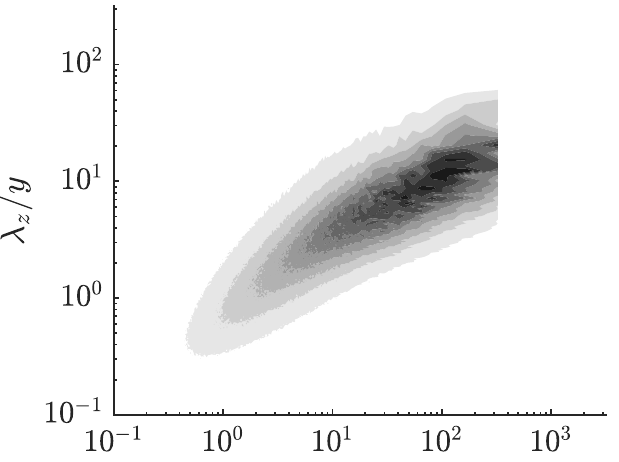}};
        \end{tikzpicture}
        \label{subfig:Phiuu2DlogDNS}}
    \sidesubfloat[]{\hspace{-0.5cm}
        \begin{tikzpicture}
            \node (img)  {\includegraphics[width = 5.0cm]{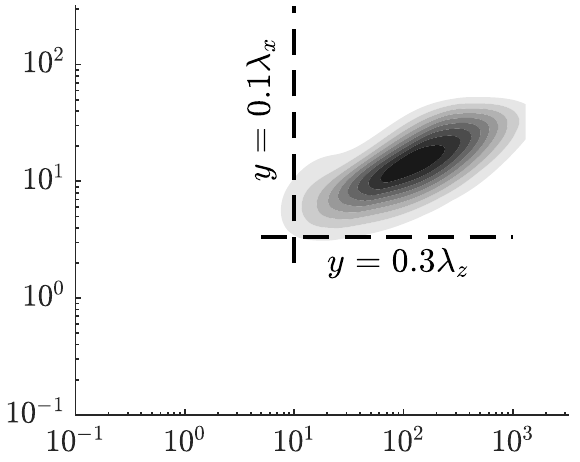}};
        \end{tikzpicture}
        \label{subfig:Phiuu2DlogDQLA}} \hfill \\
    \sidesubfloat[]{\hspace{-0.5cm}
        \begin{tikzpicture}
            \node (img)  {\includegraphics[width = 5.3cm]{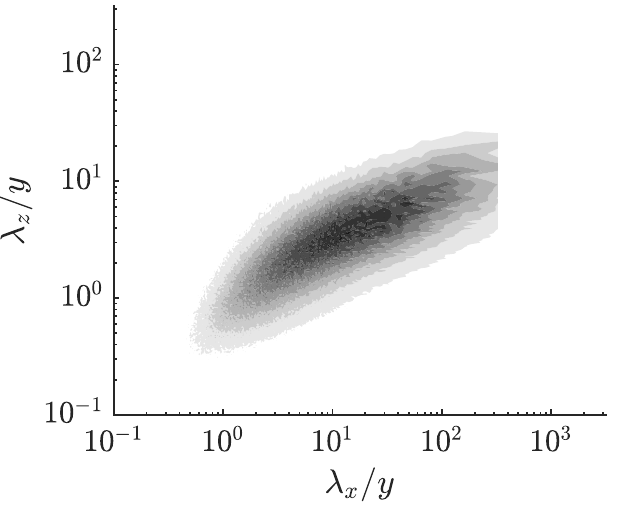}};
        \end{tikzpicture}
        \label{subfig:Phiww2DlogDNS}}
    \sidesubfloat[]{\hspace{-0.5cm}
        \begin{tikzpicture}
            \node (img)  {\includegraphics[width = 5.0cm]{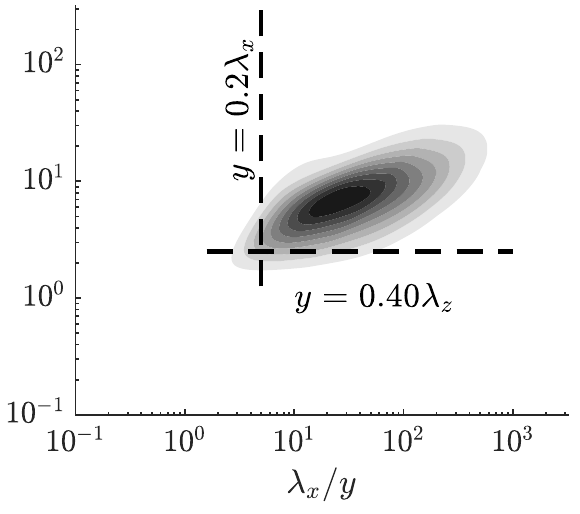}};
        \end{tikzpicture}
        \label{subfig:Phiww2DlogDQLA}}
    \end{center}
    \caption{Premultiplied two-dimensional wavenumber spectra at $y^+ \approx 400$ ($y/h\approx 0.075$) from (a,c) DNS at $Re_\tau = 5186$ and (b,d) the DQLA at $Re_\tau = 5200$: (a,b) streamwise velocity spectra; (c,d) Reynolds shear stress cospectra. The contours are normalised by 0.1 times the maximum value.}\label{fig:Phi2DlogCompare}
\end{figure}

\subsection{Two-dimensional spectra}
The two-dimensional streamwise velocity spectra and Reynolds shear stress cospectra for a fixed wall-normal location in the near-wall and logarithmic layers are shown in figures \ref{fig:kxkzPhiRe5200} and \ref{fig:Phi2DlogCompare}, respectively. The attached footprints of the energy-containing eddies from the log and outer regions are most clearly observed for the near-wall location ($y^+ \approx 15$ in figure \ref{fig:kxkzPhiRe5200}). Comparing the DNS and DQLA spectra in the near-wall region (figure \ref{fig:kxkzPhiRe5200}), there is a strong qualitative agreement, with the DQLA replicating all the features in the DNS. There is a near-wall primary peak in all the spectra, with the inactive (i.e. Reynold shear-stress absent \citep{Townsend1976}) footprints of the eddies present as an approximate linear ridge in the streamwise spectra. Consistent with the previous discussion, the attached footprint is more energetic than that present in the DNS, with the 0.30 contour describing the attached footprint in the DQLA and only the 0.10 contour in the DNS spectra. Since these features in the DQLA are modelled by the eddy viscosity diffusion operator \citep{Hwang2016MesolayerFlow,Symon2023}, this indicates a more accurate form of eddy viscosity profile in the fluctuating velocity model is required to model these features more accurately. These two-dimensional spectra also show why the peak in the streamwise rms velocity is overpredicted in the DQLA (figure \ref{fig:RMScomparison}a) -- the attached footprints of the energy-containing eddies from the log and outer regions contribute relatively more to this integrated quantity over the $k_x$-$k_z$ plane. 

The two-dimensional streamwise velocity spectra and Reynolds shear stress cospectra at a wall-normal location in the log layer are compared in figure \ref{fig:Phi2DlogCompare}. At this wall-normal location, the DNS and DQLA spectra are in good agreement for the main energy-containing features, with all of the spectra having a peak occurring at approximately the same length scales. Here, the effects of neglecting energy cascade features are most evident, with the black dashed lines on the DQLA spectra from the linear cutoff lines on the one-dimensional spectra (see figures \ref{fig:spanwiseSpectraRe5200} and \ref{fig:streamwiseSpectraKzhRe5200}). Outside these cutoff lines, there is negligible energetic content in the DQLA, whereas no simple linear cutoff was present in the DNS (figures \ref{fig:spanwiseSpectraRe5200} and \ref{fig:streamwiseSpectraKzhRe5200}). The DNS spectra are more energetic for the smaller length scales ($\lambda_x \lesssim 3y$, $\lambda_z \lesssim 3y$), which are associated with energy cascade to small scales for dissipation. The contribution of these energy cascade features to the turbulence intensities is expected to be reasonably small. 

 \section{Scaling behaviour up to $\boldsymbol{Re_\tau = 10^5}$}\label{sec:DDQLAscaling}
 \begin{figure}
    \begin{center}
    \sidesubfloat[]{\hspace{-0.5cm}
        \begin{tikzpicture}
            \node (img)  {\includegraphics[width = 5.2cm]{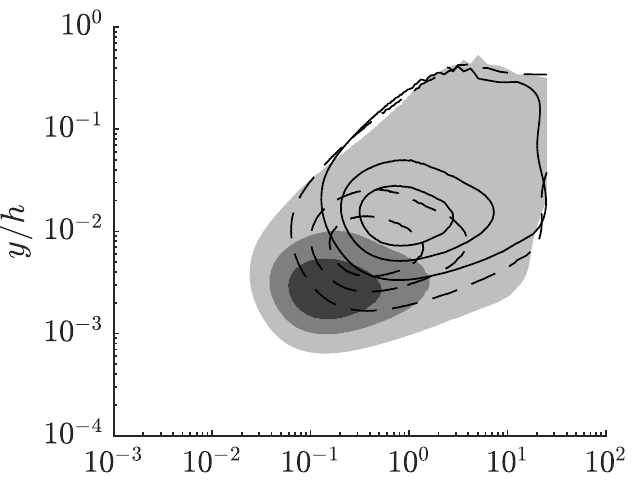}};
        \end{tikzpicture}
        \label{subfig:OuterDNSkxPhiUU}}
    \sidesubfloat[]{
        \begin{tikzpicture}
            \node (img)  {\includegraphics[width = 4.9cm]{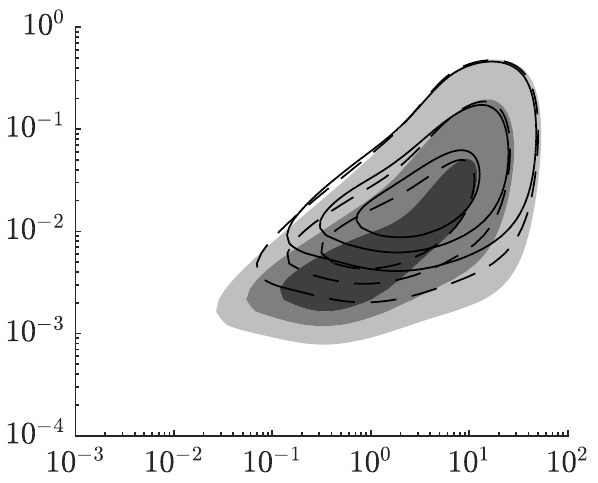}};
        \end{tikzpicture}
        \label{subfig:OuterQLAkxPhiUU}} \hfill \\
    \sidesubfloat[]{\hspace{-0.5cm}
        \begin{tikzpicture}
            \node (img)  {\includegraphics[width = 5.2cm]{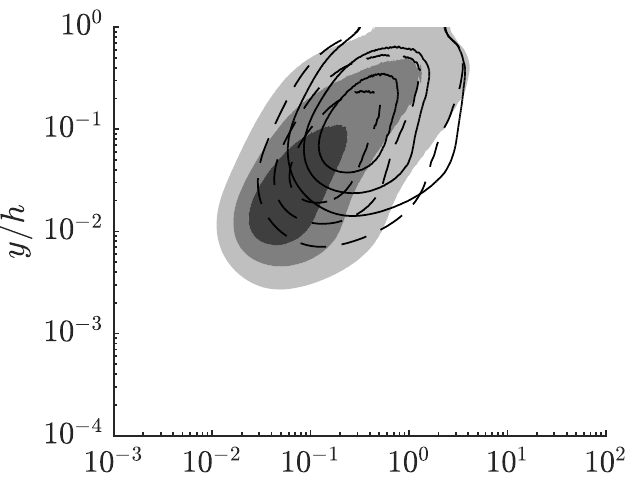}};
        \end{tikzpicture}
        \label{subfig:OuterDNSkxPhiVV}}
    \sidesubfloat[]{
        \begin{tikzpicture}
            \node (img)  {\includegraphics[width = 4.9cm]{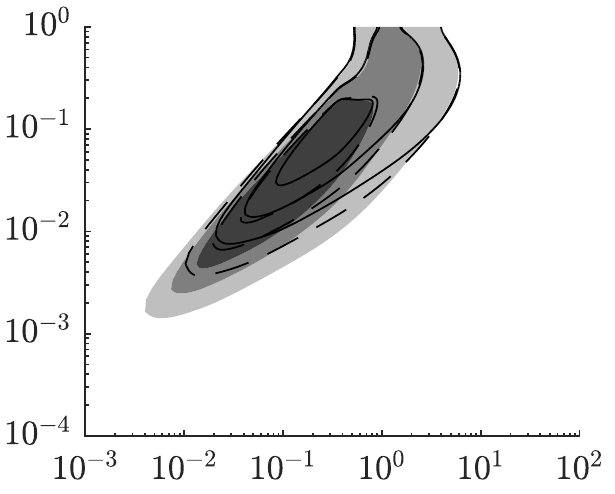}};
        \end{tikzpicture}
        \label{subfig:OuterQLAkxPhiVV}} \hfill \\
    \sidesubfloat[]{\hspace{-0.5cm}
        \begin{tikzpicture}
            \node (img)  {\includegraphics[width = 5.2cm]{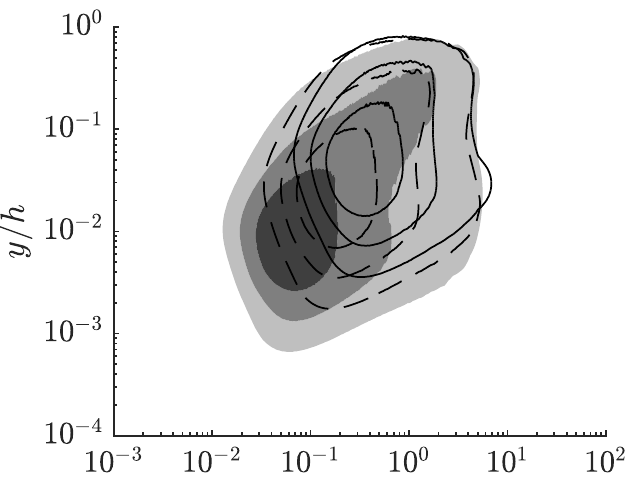}};
        \end{tikzpicture}
        \label{subfig:OuterDNSkxPhiWW}}
    \sidesubfloat[]{
        \begin{tikzpicture}
            \node (img)  {\includegraphics[width = 4.9cm]{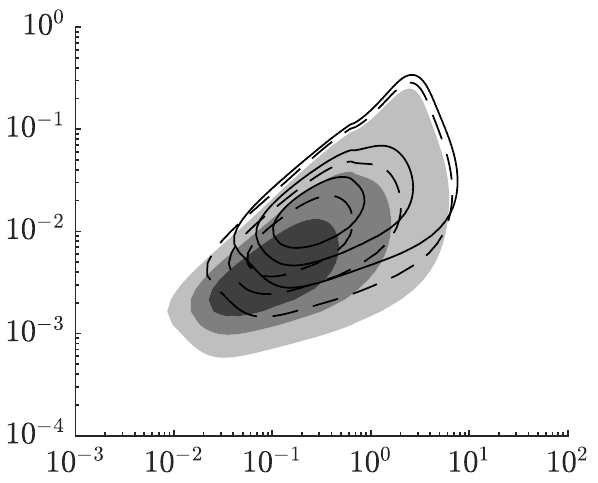}};
        \end{tikzpicture}
        \label{subfig:OuterQLAkxPhiWW}} \hfill \\
    \sidesubfloat[]{\hspace{-0.5cm}
        \begin{tikzpicture}
            \node (img)  {\includegraphics[width = 5.2cm]{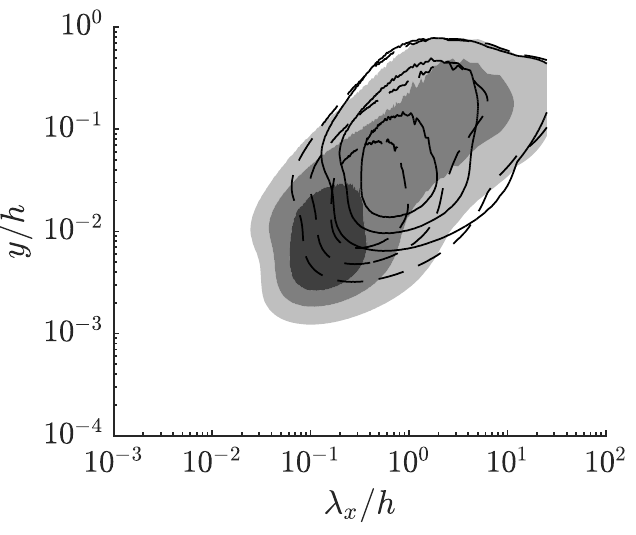}};
        \end{tikzpicture}
        \label{subfig:OuterDNSkxPhiUV}}
    \sidesubfloat[]{
        \begin{tikzpicture}
            \node (img)  {\includegraphics[width = 4.9cm]{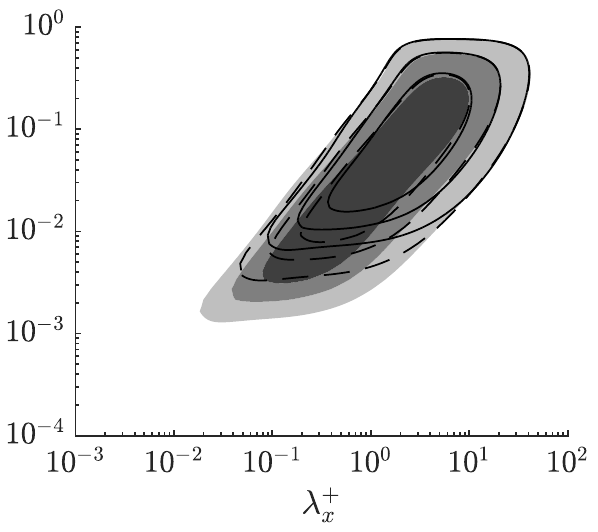}};
        \end{tikzpicture}
        \label{subfig:OuterQLAkxPhiUV}}
    \end{center}
    \caption{Outer-scaled streamwise one-dimensional spectra from (a,c,e,g) DNS \citep{Lee2015} and (b,d,f,h) the DQLA: (a,b) streamwise velocity; (c,d) wall-normal velocity; (e,f) spanwise velocity; (g,h) Reynolds shear stress. Here $Re_\tau \approx 5200, 2000, 1000$ for the shaded, dashed and solid line contours, respectively. The contour levels are chosen to be 0.25, 0.50 and 0.75 times the maximum value for comparison, except in (d) where all contours levels are given by 0.25, 0.50 and 0.75 times the maximum value of the $Re_\tau = 5200$ spectra.}\label{fig:OuterkxPhi}
\end{figure}

\begin{figure}
    \begin{center}
    \sidesubfloat[]{\hspace{-0.5cm}
        \begin{tikzpicture}
            \node (img)  {\includegraphics[width = 5.3cm]{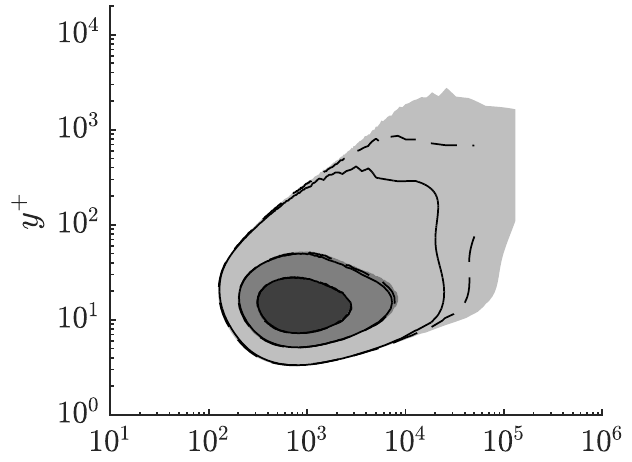}};
        \end{tikzpicture}
        \label{subfig:PhiuuKzh14}}
    \sidesubfloat[]{
        \begin{tikzpicture}
            \node (img)  {\includegraphics[width = 4.9cm]{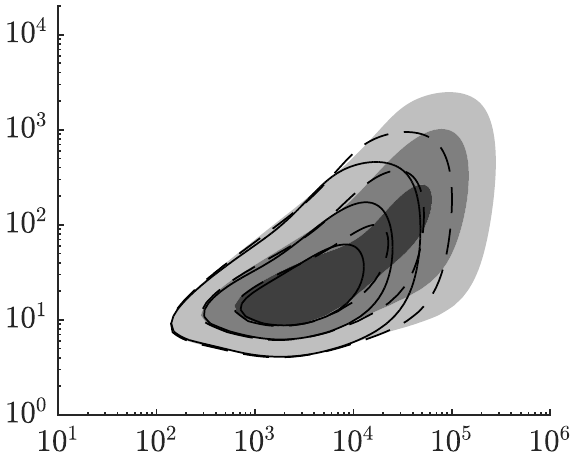}};
        \end{tikzpicture}
        \label{subfig:Innerkxphi_uuDNS}} \hfill \\
    \sidesubfloat[]{\hspace{-0.5cm}
        \begin{tikzpicture}
            \node (img)  {\includegraphics[width = 5.3cm]{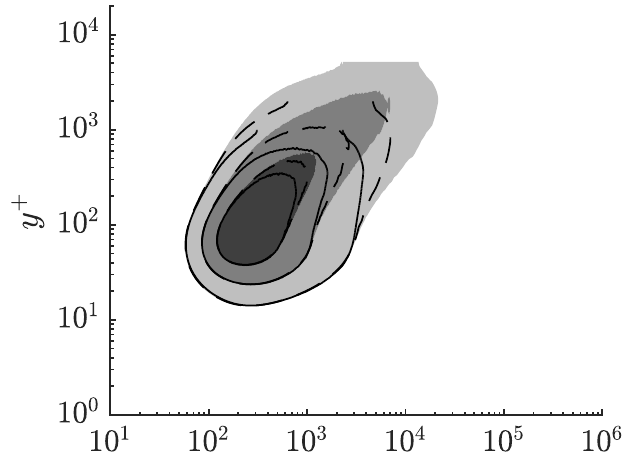}};
        \end{tikzpicture}
        \label{subfig:Innerkxphi_vvDNS}}
    \sidesubfloat[]{
        \begin{tikzpicture}
            \node (img)  {\includegraphics[width = 4.9cm]{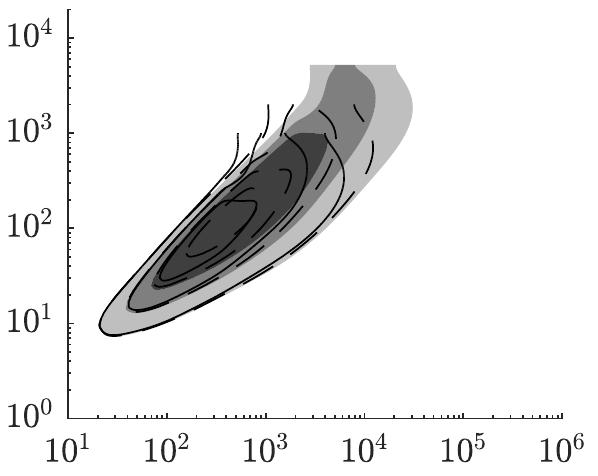}};
        \end{tikzpicture}
        \label{subfig:Innerkxphi_vvQLA}} \hfill \\
    \sidesubfloat[]{\hspace{-0.5cm}
        \begin{tikzpicture}
            \node (img)  {\includegraphics[width = 5.3cm]{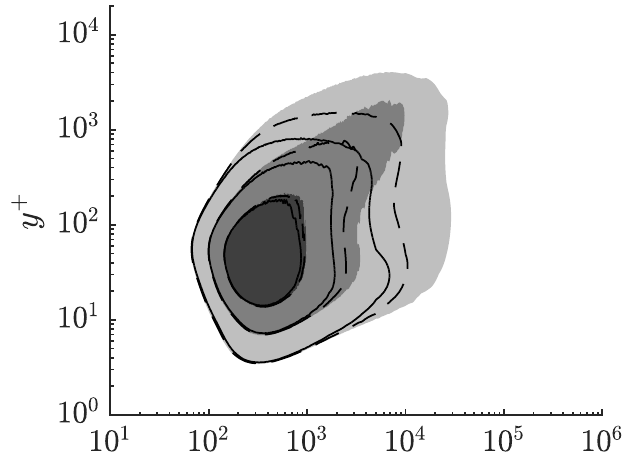}};
        \end{tikzpicture}
        \label{subfig:Innerkxphi_wwDNS}}
    \sidesubfloat[]{
        \begin{tikzpicture}
            \node (img)  {\includegraphics[width = 4.9cm]{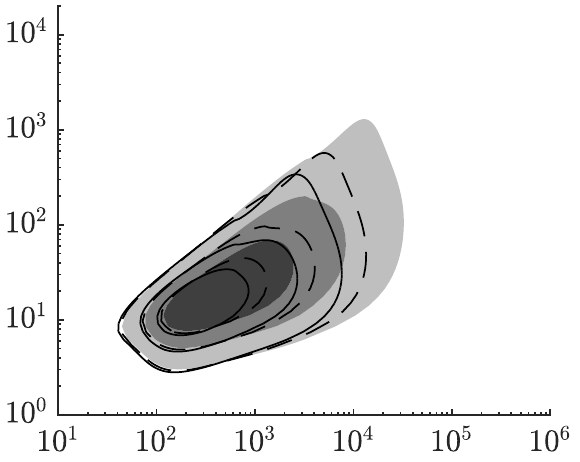}};
        \end{tikzpicture}
        \label{subfig:Innerkxphi_wwQLA}} \hfill \\
    \sidesubfloat[]{\hspace{-0.5cm}
        \begin{tikzpicture}
            \node (img)  {\includegraphics[width = 5.3cm]{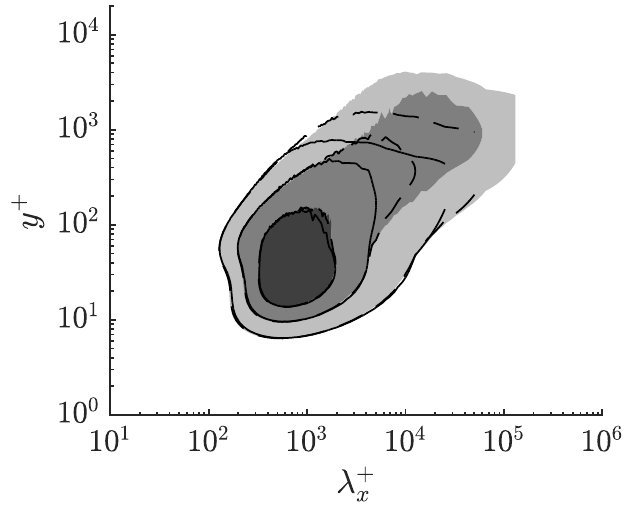}};
        \end{tikzpicture}
        \label{subfig:Innerkxphi_uvDNS}}
    \sidesubfloat[]{
        \begin{tikzpicture}
            \node (img)  {\includegraphics[width = 4.9cm]{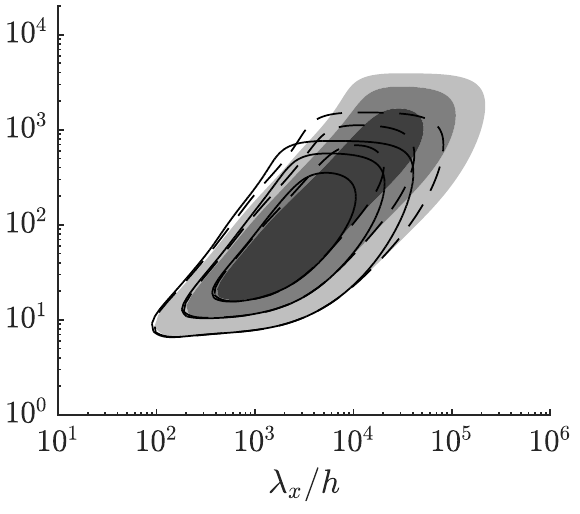}};
        \end{tikzpicture}
        \label{subfig:Innerkxphi_uvQLA}}
    \end{center}
    \caption{Inner-scaled streamwise one-dimensional spectra from (a,c,e,g) DNS \citep{Lee2015} and (b,d,f,h) the DQLA: (a,b) streamwise velocity; (c,d) wall-normal velocity; (e,f) spanwise velocity; (g,h) Reynolds shear stress. Here $Re_\tau \approx 5200, 2000, 1000$ for the shaded, dashed and solid line contours, respectively. The contour levels are chosen to be 0.25, 0.50 and 0.75 times the maximum value for comparison, except in (d) where all contours levels are given by 0.25, 0.50 and 0.75 times the maximum value of the $Re_\tau = 5200$ spectra.}\label{fig:InnerkxPhi}
\end{figure}

\subsection{Spectra}
The DQLA built from the DNS data at $Re_\tau \approx 5200$ is now repeated for Reynolds numbers ranging from $Re_\tau = 10^3$ to  $Re_\tau = 10^5$, given its modelling scope to extrapolate to other Reynolds numbers. Using the self-similar weight $W_{k_x}(k_x/k_z)$ constructed with the DNS data at $Re_\tau \approx 5200$, the DQLA is performed by solving (\ref{eq:DDQLA}) at the range of the Reynolds numbers considered. Firstly, the Reynolds scaling of the one-dimensional spectra is compared between the DNS and DQLA for $Re_\tau \approx 1000, 2000, 5200$. The qualitative scaling behaviour in the DQLA is found to be identical to the MQLA in the spanwise one-dimensional spectra and is not displayed here \cite[see][for a more complete discussion]{Hwang2020,Skouloudis2021ScalingApproximation}. There is a strong qualitative agreement between them, with the DQLA and DNS exhibiting inner-scaling features for $\mathcal{O}(10) \lesssim \lambda_z^+ \lesssim \mathcal{O}(10^3) $ and outer-scaling behaviour for $\lambda_z \approx \mathcal{O}(h)$. The attached footprints in the wall-parallel velocity spectra from the DQLA importantly exhibit inner-scaling behaviour \citep{Hwang2016MesolayerFlow}. Overall the spanwise velocity spectra are consistent with the attached eddy hypothesis, with qualitative corrections due to the incorporation of viscous effects at a finite Reynolds number \citep{Skouloudis2021ScalingApproximation}.

The outer- and inner-scaled streamwise one-dimensional spectra are shown in figures \ref{fig:OuterkxPhi} and \ref{fig:InnerkxPhi}, respectively. Note that, due to the peak in the wall-normal velocity spectra occurring at logarithmic streamwise length scales in the DQLA, the contour levels for the inner-scaled streamwise one-dimensional spectra for the wall-normal velocity (figure \ref{fig:InnerkxPhi}d) are chosen to follow absolute values rather than normalised ones to exhibit the inner-scaling behaviour. Despite the differences between the spectra at a single Reynolds number, as described in \S\ref{sec:DDQLAre5200}, the scaling behaviour in the DQLA compares very well with the DNS. All the spectra are energetic, spanning from $\lambda_x^+ = \mathcal{O}(10^2)$ up to $\lambda_x/h = \mathcal{O}(10)$. Both the velocity and Reynolds shear stress spectra have an inner-scaling near-wall peak, with the outer part of the spectra scaling well in outer units. Like the spanwise one-dimensional spectra \citep{Hwang2020,Skouloudis2021ScalingApproximation}, the wall-parallel velocity spectra have an attached footprint, scaling well in outer and inner units. 


\begin{figure}
    \begin{center}
    \sidesubfloat[]{\hspace{-0.5cm}
        \begin{tikzpicture}
            \node (img)  {\includegraphics[width = 5.1cm]{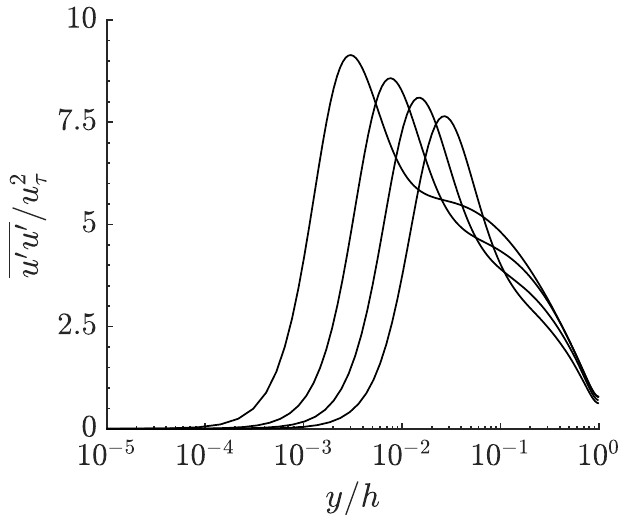}};
            \draw[yshift=-1.2cm,xshift=0.7cm,->] (0:0) -- (150:1.9) node[right] {$Re_\tau$};
        \end{tikzpicture}
        \label{subfig:DNSouterUU}}
    \sidesubfloat[]{\hspace{-0.5cm}
        \begin{tikzpicture}
            \node (img)  {\includegraphics[width = 5.0cm]{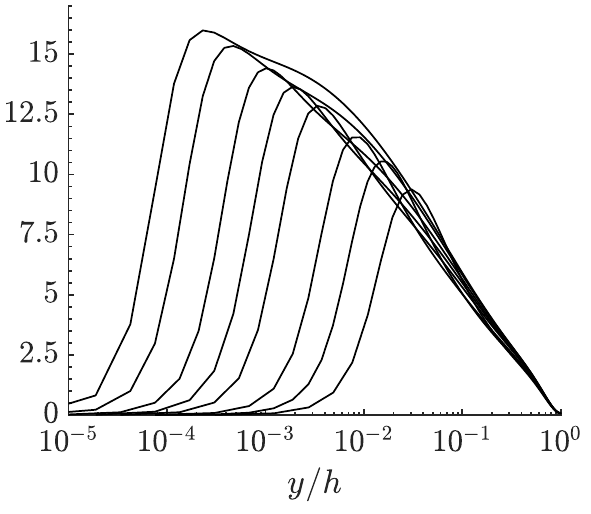}};
            \draw[yshift=-1.2cm,xshift=0.6cm,->] (0:0) -- (150:2.5) node[right] {$Re_\tau$};
        \end{tikzpicture}
        \label{subfig:QLAouterUU}} \hfill \\
    \sidesubfloat[]{\hspace{-0.5cm}
        \begin{tikzpicture}
            \node (img)  {\includegraphics[width = 5.0cm]{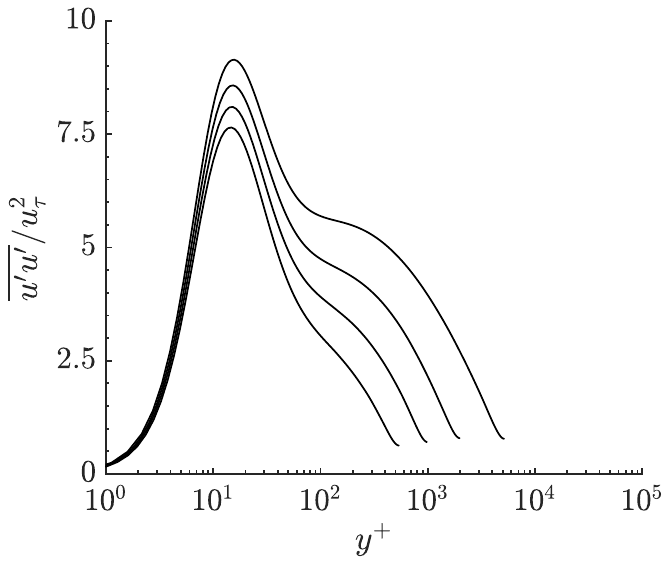}};
            \draw[yshift=-1.2cm,xshift=-0.3cm,->] (0:0) -- (30:2.0) node[right] {$Re_\tau$};
        \end{tikzpicture}
        \label{subfig:DNSinnerUU}}
    \sidesubfloat[]{\hspace{-0.5cm}
        \begin{tikzpicture}
            \node (img)  {\includegraphics[width = 5.0cm]{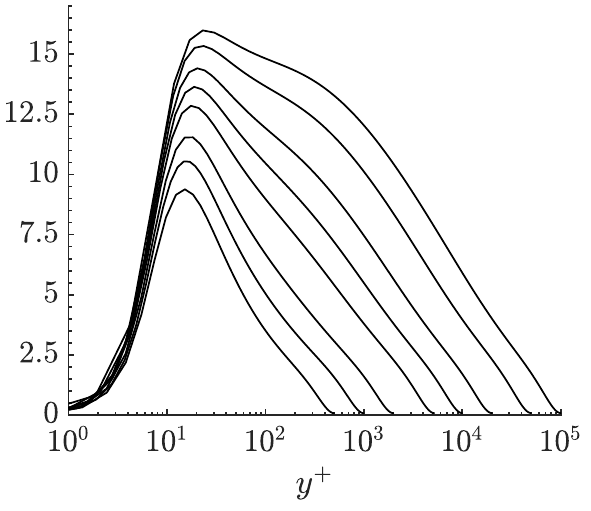}};
            \draw[yshift=-1.2cm,xshift=-0.3cm,->] (0:0) -- (30:2.5) node[right] {$Re_\tau$};
        \end{tikzpicture}
        \label{subfig:QLAinnerUU}}
    \end{center}
    \caption{Streamwise turbulence intensity profiles from (a,c) DNS \citep{Lee2015} and (b,d) the DQLA in outer-scaled coordinates. Here $Re_\tau = 550, 1000, 1994, 5185$ for DNS and $Re_\tau = 500, 1000, 2000, 5200, 10000, 20000, 50000, 100000$ for the DQLA. }\label{fig:InnerOuterUUintense}
\end{figure}

\subsection{Turbulence intensity}
The predictive capabilities of the DQLA are now used up to $Re_\tau = 10^5$, with the focus on the streamwise turbulence intensity profiles. The other components are consistent with the MQLA \citep{Hwang2020}, albeit with a reduced level of anisotropy, as presented in \S\ref{sec:DDQLAre5200}. The profiles are plotted in the inner- and outer-scaled coordinates in figure \ref{fig:InnerOuterUUintense}. The scaling behaviour of the streamwise intensity profiles in DNS and DQLA share the same key features: a near-wall peak at $y^+\approx 15$ at relatively low Reynolds numbers ($Re_\tau \lesssim 5000$) and an approximate logarithmic decay when scaled in outer units. This behaviour is more evident in the DQLA for $Re_\tau \gtrsim 5200$. For these larger Reynolds numbers, the streamwise intensity profile is consistent with \cite{hwang_hutchins_marusic_2022}, in which the spectrum-based attached eddy model of \citep{Perry1986ATurbulence} was extended for finite Reynolds numbers with an experimental data of \citep{Samie2018}. In the upper logarithmic layer (or inertial sublayer) from $y^+ = 3.6Re_\tau^{0.5}$ up to $y/h = 0.2$, this model yields the following form of streamwise turbulence intensity
\begin{subequations}
\begin{equation}\label{eq:turinten}
    \frac{\overline{u'u'}}{u_\tau^2} = -A(Re_\tau)\ln(y/h) + B(Re_\tau),
\end{equation}
where 
\begin{equation}\label{eq:viscousTownsendPerry}
    A(Re_\tau) = A_0 + A_1(Re_\tau).
\end{equation}
\end{subequations}
Here, $A(Re_\tau)$ and $B(Re_\tau)$ are supposed to be constants in the limit of $Re_\tau$, and they vary slowly with $Re_\tau$ at finite $Re_\tau$ \citep{hwang_hutchins_marusic_2022}.

\begin{figure}
    \begin{center}
    \sidesubfloat[]{\hspace{-0.6cm}
        \begin{tikzpicture}
            \node (img)  {\includegraphics[width = 4.5cm]{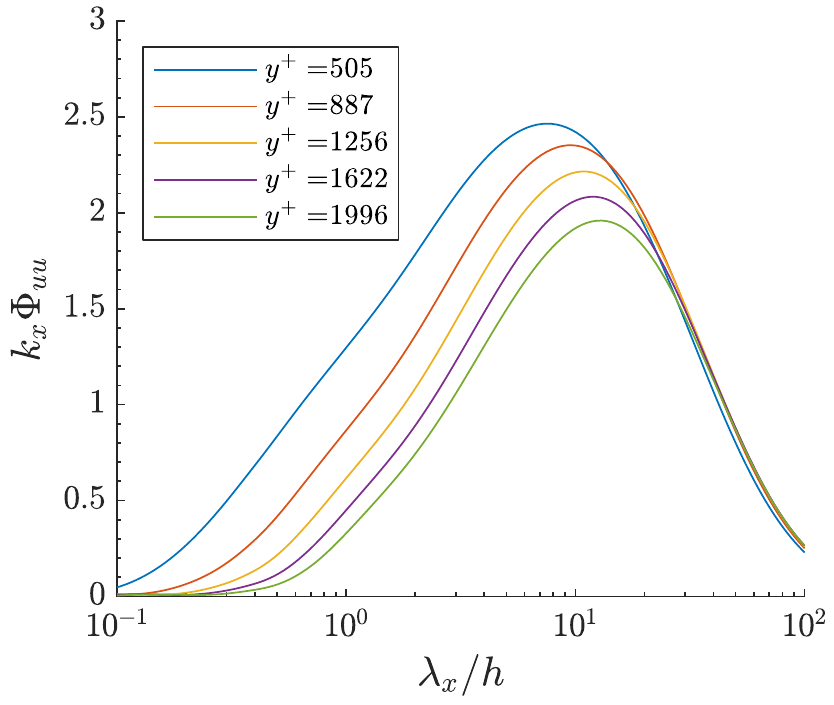}};
        \end{tikzpicture}
        \label{subfig:innerSpecScaling}}\hspace{-0.6cm}
    \sidesubfloat[]{\hspace{-0.6cm}
        \begin{tikzpicture}
          \node (img)  {\includegraphics[width = 4.5cm]{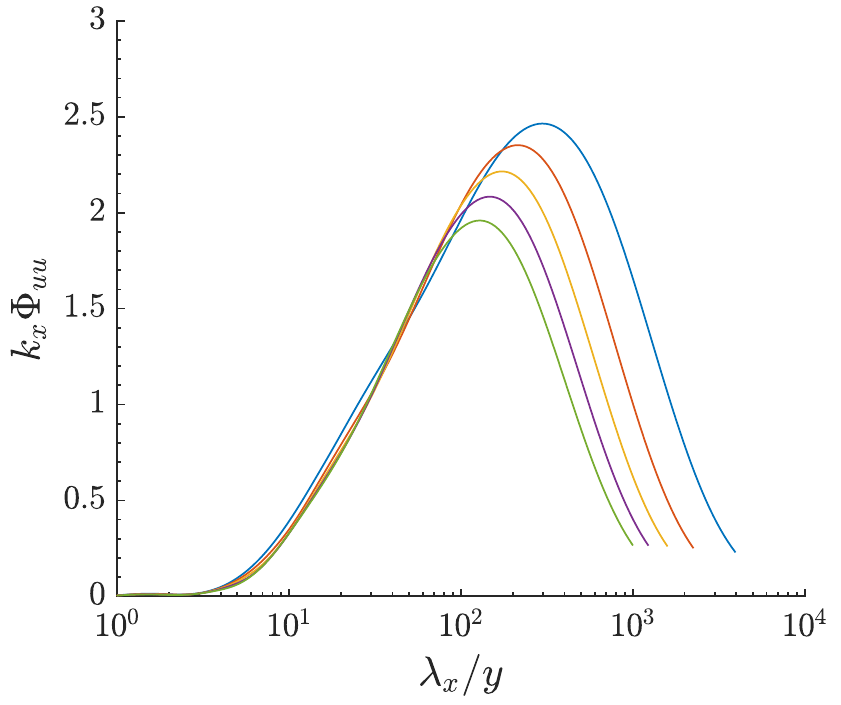}};
        \end{tikzpicture} \hspace{-0.6cm}
        \label{subfig:outerSpecScaling}}
        \sidesubfloat[]{\hspace{-0.5cm}
        \begin{tikzpicture}
          \node (img)  {\includegraphics[width = 4.4cm]{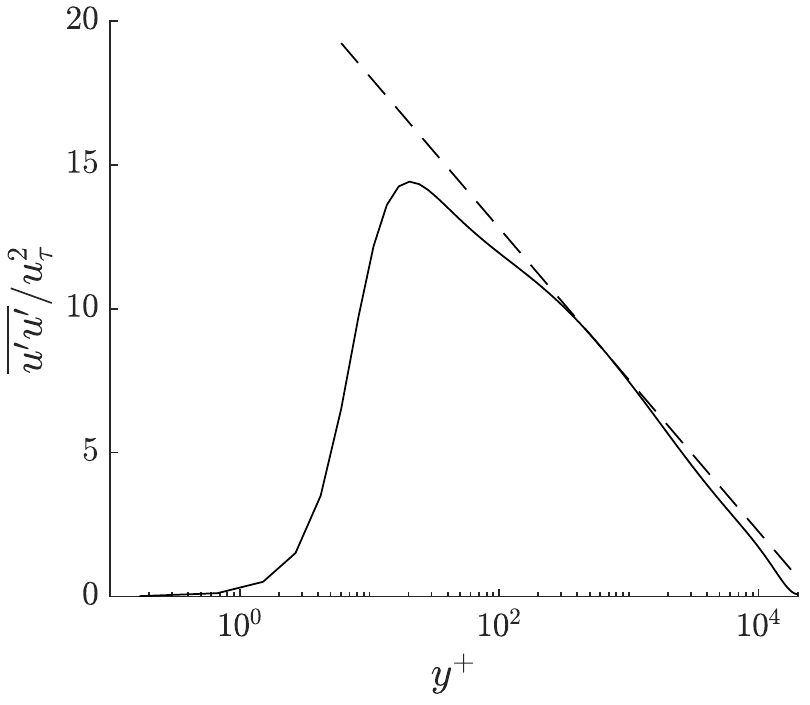}};
        \end{tikzpicture}
        \label{subfig:uuFitExample}}
    \end{center}
    \caption{Premultiplied streamwise one-dimensional spectra at various wall-normal locations for $Re_\tau = 20000$ in $(a)$ outer scaling coordinates $\lambda_x/h$ and $(b)$ logarithmic coordinates $\lambda_x/y$ and $(c)$ the streamwise turbulence intensity profile (solid) with the attached eddy hypothesis approximation following \cite{hwang_hutchins_marusic_2022}.  } \label{fig:nearWallSpectraScaling}
\end{figure}

\begin{table}
    \centering
    \begin{tabular}{c c c c c c c} \hline
         $Re_\tau$ $\times 10^{-4}$ & & 0.52 & 1 & 2 & 5 & 10 \\ \hline
         $A(Re_\tau)$ & & 2.12 & 2.21 & 2.31 & 2.41 & 2.47 \\ 
         $B(Re_\tau)$ & & 0.61 & 0.64 & 0.67 & 0.69 & 0.71 \\ \hline
    \end{tabular}
    \caption{The Reynolds-number dependent model constants for the streamwise turbulence intensity determined following \cite{hwang_hutchins_marusic_2022}.}
    \label{tab:TownsendPerryConstant}
\end{table}

An essential prerequisite of the model in \cite{hwang_hutchins_marusic_2022} is the existence of $y$- and $h$-scaling regions of one-dimensional spectra in the upper logarithmic layer. Figure \ref{fig:nearWallSpectraScaling} (a,b) show that the DQLA successfully reproduces such spectra in the upper logarithmic layer \cite[compare with figures 3(a,b) in][]{hwang_hutchins_marusic_2022} like the experimental data of \cite{Samie2018}. Following \cite{hwang_hutchins_marusic_2022}, $A(Re_\tau)$ and $B(Re_\tau)$ in (\ref{eq:turinten}) are subsequently approximated from the spectra at all Reynolds numbers: 
\begin{subequations}\label{eq:viscousAEH}
\begin{equation}
        A(Re_\tau) = \left[\ln\left(\frac{a_{x,u}}{y}\right) - \ln\left(\frac{a_{x,l}}{h}\right) \right]^{-1}\int_{\ln\left(a_{x,l}/h\right)}^{\ln\left(a_{x,u}/y\right)}\frac{k_x\Phi_{uu}(k_x,y/h)}{u_\tau^2}\mathrm{d}\ln(k_x),
\end{equation}
\begin{equation}
        B(Re_\tau) = A(Re_\tau)\ln\left(\frac{a_{x,u}}{a_{x,l}}\right).
\end{equation}
\end{subequations}
Here, $a_{x,u}$ and $a_{x,l}$ are dimensionless constants associated with the upper and lower limits of the integration of the streamwise spectra. However, they have to be chosen through inspection of figures \ref{fig:nearWallSpectraScaling}(a,b), they must be `constants' for all Reynolds numbers. The approximation \eqref{eq:viscousAEH} reduces down to using the mean-value-theorem to approximate the spectra across these upper and lower limits, with the mean value, or in this case, the Townsend-Perry constant $A(Re_\tau)$, consisting of a universal component $A_0$ and a viscous correction $A_1(Re_\tau)$. Hence the upper and lower limits lie at values where the spectra scale in outer ($\lambda_x/h$) and logarithmic ($\lambda_x/y$) coordinates,  respectively. After trial and improvement of the fitting procedure, the upper and lower limits are set with $a_{x,u} = \pi$ and $a_{x,l} = 4\pi/3$, with the comparison of the approximation and streamwise turbulence intensity profile shown in figure \ref{fig:nearWallSpectraScaling}(c).

Table \ref{tab:TownsendPerryConstant} reports the values of $A(Re_\tau)$ and $B(Re_\tau)$ obtained. Consistent with the growing trend of $A(Re_\tau)$ and $B(Re_\tau)$ observed in \cite{hwang_hutchins_marusic_2022} with the experimental data from \cite{Samie2018}, their values obtained from the DQLA data also slowly grow. Importantly, their growth rate tends to be smaller on increasing $Re_\tau$ from Table \ref{tab:TownsendPerryConstant}, indicating that they would reach constant values. This trend is consistent with the theoretical model of \cite{hwang_hutchins_marusic_2022}, which becomes identical to the classical attached eddy model in the limit of $Re_\tau \rightarrow \infty$ \citep{Townsend1976,Perry1982OnTurbulence,Perry1986ATurbulence}. 


Aside from the agreement between the near-wall peak and logarithmic decay, the DQLA still does not have a clear plateau behaviour for $y^+ \approx 200$, although one starts to emerge for $Re_\tau \gtrsim 20000$. This is likely due to the overly energetic response of the large-scale motions present in the current model, as discussed in \S\ref{sec:DDQLAre5200}. In the DQLA, the primary near-wall peaks are much less distinct than those in the DNS, with the outer-scaling parts of the spectra and their attached features remaining relatively more energetic in the spectra. 

\begin{figure}
    \begin{center}
    \sidesubfloat[]{\hspace{-0.5cm}
        \begin{tikzpicture}
            \node (img)  {\includegraphics[width = 5.0cm]{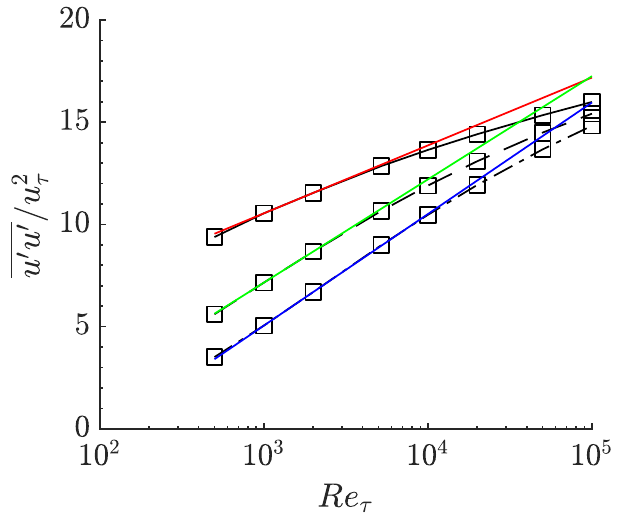}};
        \end{tikzpicture}
        \label{subfig:peakReScaling}}\hspace{1cm}
    \sidesubfloat[]{\hspace{-0.5cm}
        \begin{tikzpicture}
          \node (img)  {\includegraphics[width = 5.2cm]{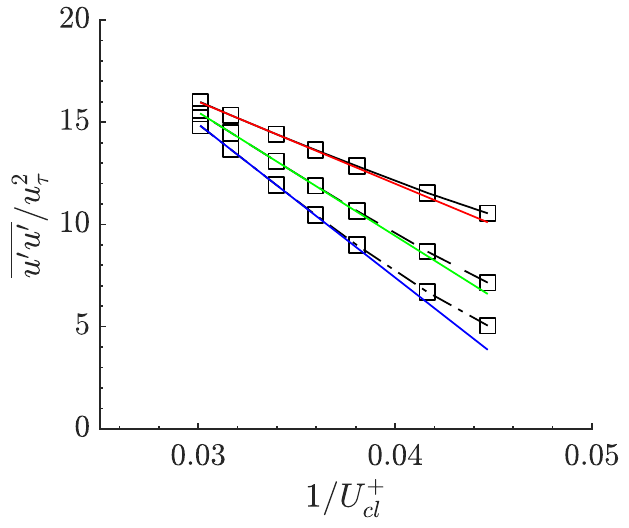}};
        \end{tikzpicture}
        \label{subfig:uCl_scaling}}
    \end{center}
    \caption{The Reynolds scaling behaviour of the streamwise turbulence intensity based on (a) $\log Re_\tau$ and (b) inner-scaled centreline velocity $U_{cl}^+$. The wall-normal locations correspond to the peak (solid); $y^+ = 50$ (dashed); $y^+ = 100$ (dash-dotted). The coloured lines correspond to (a) $\overline{u'u'}/u_\tau^2 = a_1 + b_1\ln(Re_\tau)$ fitted between $Re_\tau = 1000-2000$ and (b) $\overline{u'u'}/u_\tau^2 = a_2 + b_2/U_{cl}^+$ fitted between $Re_\tau = 20000-50000$.} \label{fig:nearWallScaling}
\end{figure}

Finally, figure \ref{fig:nearWallScaling} shows the scaling behaviour of the near-wall peak and two other inner-scaling locations, $y^+= 50,100$. Like the MQLA \citep{Hwang2020,Skouloudis2021ScalingApproximation} in figure\ref{fig:nearWallScaling} (a), the DQLA shows the deviation of near-wall intensities from the classical logarithmic scaling predicted by an extension of the original attached eddy model \citep{Marusic2003}: i.e. $\overline{u'u'}/u_\tau^2 \sim \ln Re_\tau$. 
Instead, consistent with the recent findings from a variant of the MQLA \citep{Skouloudis2021ScalingApproximation}, the near-wall streamwise intensities are inversely proportional to the inner-scaled centreline velocity, with the coloured lines given by fits of 
\begin{equation}\label{eq:scaling}
    \frac{\overline{u'u'}}{u_\tau^2} = C - D/U_{cl}^+,
\end{equation}
favoring the prediction made by \cite{Monkewitz2015} for a turbulent boundary layer using an asymptotic expansion of near-wall turbulence statistics. Notably, the MQLA \citep{Skouloudis2021ScalingApproximation} and DQLA provided the same scaling associated with $1/U_{cl}^+$. However, their construction is quite different, especially in that of the full velocity spectra. This suggests that the scaling behaviour of (\ref{eq:scaling}) is inherently from the linearised Navier-Stokes equations in (\ref{subeq:fluctuatingVelocity}) with the model nonlinear term (\ref{subeq:modelNLterm}) rather than a peculiar feature emerging from the construction of the quasi-linear approximation. It also strongly indicates that there must exist a mathematical structure underpinning (\ref{eq:scaling}). Given that the DQLA successfully replicates the scaling behaviour of the streamwise velocity in the upper logarithmic layer, this result should not be lightly considered. Importantly, there is growing evidence that the scaling of $\overline{u'u'}/u_\tau^2 \sim \ln Re_\tau$ from the classical attached eddy model may not be valid, as the original attached eddy model is built by ignoring the viscous effect in the near-wall region. In this respect, it is worth mentioning an alternative proposed by the recent work by \cite{Chen2021}, where a scaling proportional to $Re_\tau^{-1/4}$ was proposed instead of $1/U_{cl}^+$ scaling. Such a scaling may fit well with the near-wall streamwise turbulence intensity of DQLA. However, it was recently shown that, in practice, only a negligibly small difference has been found between the $1/U_{cl}^+$ and $Re_\tau^{-1/4}$ scalings upon increasing the Reynolds number \citep{Nagib2022}. To the best of the authors' knowledge, the correct scaling behaviour of the near-wall streamwise turbulence intensity is currently an issue of debate. One of the authors of the present study \citep{Hwang2022} makes an 
on-going effort to address this issue thoroughly in the future.

\section{Summary}\label{sec:conclusions}
The MQLA \citep{Hwang2020} has been extended in the present study by including streamwise variations of turbulence spectra. To extend the MQLA while still maintaining its predictive nature, self-similarity was used to determine the statistical structure of the forcing. By using the universal nature and growing significance of the logarithmic layer on increasing Reynolds number, a set of self-similar weights were determined by matching the two-dimensional spectra with respect to the streamwise wavenumber and wall-normal location at a single spanwise wavenumber of the linearised Navier-Stokes equations to those of a DNS performed at $Re_\tau \approx 5200$. By reconstructing the velocity spectra from the leading POD modes of the linearised Navier-Stokes equations, the two-dimensional spectra generated reasonably well replicated the DNS spectra. In doing so, the energy cascade-associated features in the spectra were neglected, in line with the attached eddy hypothesis, for the model to be extrapolatable to other Reynolds numbers. From this self-similar weighting with respect to the streamwise wavenumber, the self-consistent determination of the Reynolds shear stress was implemented following \cite{Hwang2020}, completing the DQLA framework. 

The DQLA allows complete determination of the two-dimensional velocity spectra and all subsequent statistics, with results compared between the MQLA, DQLA and DNS. It was shown that the DQLA offers significant quantitative improvements compared to the MQLA. In particular, it significantly reduces the anisotropy in the turbulence intensities while providing the streamwise wavenumber spectra, the scaling of which is consistent with that of DNS. While the DQLA did improve turbulence statistics and spectra compared to the MQLA, there were still some qualitative differences between the DQLA and DNS results. This was demonstrated most clearly in the streamwise one-dimensional spectra, the intensity of which was much stronger than that of the DNS in the region close to the wall while lacking the energetic content at length scales associated with the streak instability and/or transient growth \cite[][]{Schoppa2002CoherentTurbulence,deGiovanetti2017StreakMotions,Lozano-Duran2021Cause-and-effectTurbulence}. Aside from the qualitative differences in the spectra, the DQLA framework was shown to retain the predictive capabilities of the MQLA with the scaling behaviour of turbulence intensities and spectra in qualitative agreement with the DNS. In particular, it offers the scaling behaviour consistent with the recent theoretical model of \cite{hwang_hutchins_marusic_2022}, where the spectrum-based attached eddy model in \cite{Perry1986ATurbulence} was extended for finite Reynolds numbers. Also, like the MQLA, the near-wall peak turbulence intensity was inversely proportional to the inner-scaled scaled centreline mean velocity, deviating from the classical prediction based on the attached eddy model. 

\section*{Acknowledgements}
J.H. is supported by Doctoral Training Partnership (DTP) scholarship from the Engineering Physical Science Research Council (EPSRC) in the UK. Y.H. gratefully acknowledges the support of the Leverhulme trust (RPG-123-2019) and EPSRC (EP/T009365/1).

\section*{Declaration of interest}
The authors report no conflict of interest.

\appendix 
\section{Sensitivity of optimisation procedure}\label{sec:sensOfWeight}
\begin{figure}[h!]
    \begin{center}
        \begin{tikzpicture}
            \node (img)  {\includegraphics[width = 5.0cm]{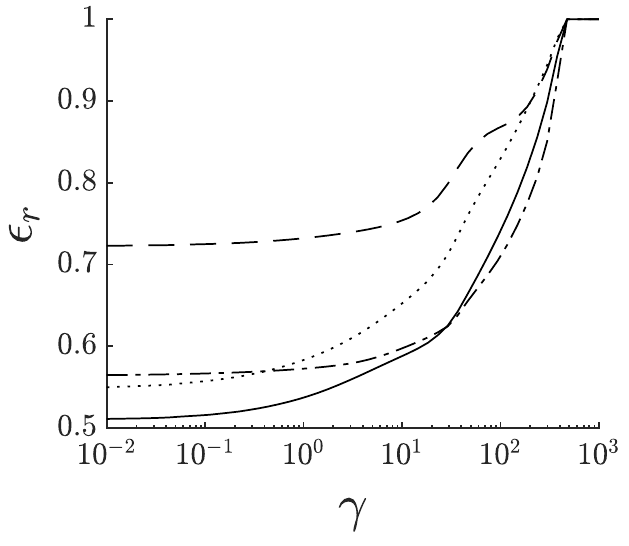}};
        \end{tikzpicture}
        \label{subfig:gammaErrorTrace}
    \end{center}
    \caption{The trade off curve between the componentwise errors in \eqref{eq:NormMin}, where $\epsilon_r = {||\Phi_{r}^{\mathrm{DNS}}-\Phi_{r}||_Q}/{||\Phi_{r}^{\mathrm{DNS}}||_Q}$ for $k_zh = 14$ for the streamwise (solid), wall-normal (chain), spanwise velocity spectra (dotted)  and Reynolds shear-stress cospectra  (dashed).}\label{fig:gammaTrace}
\end{figure}

To select an appropriate value for $\gamma$, figure \ref{fig:gammaTrace} shows a trade-off curve between the regularisation parameter $\gamma$ and the errors with respect to the $Q-$norm for each of the spectra. As the errors in all of the components are approximately monotonic, the weights were determined by setting $\gamma = 0.5$ and using trial and inspection in varying $\gamma$ until the streamwise weights and the velocity spectra are sufficiently smooth and in good qualitative agreement with the DNS velocity spectra.

\begin{figure}
    \begin{center}
    \sidesubfloat[]{\hspace{-0.5cm}
        \begin{tikzpicture}
            \node (img)  {\includegraphics[width = 5.0cm]{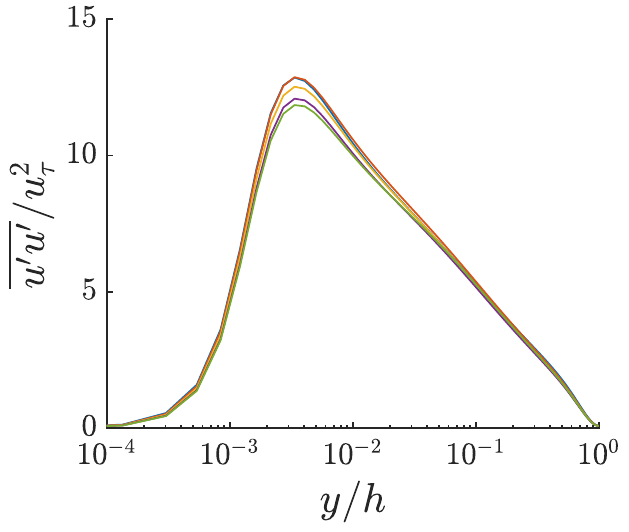}};
        \end{tikzpicture}
        \label{subfig:WkxSensUU}}
    \sidesubfloat[]{
        \begin{tikzpicture}
            \node (img)  {\includegraphics[width = 5.2cm]{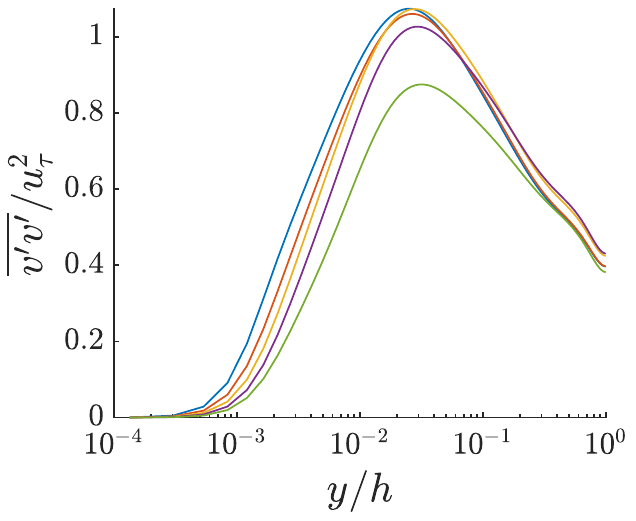}};
        \end{tikzpicture}
        \label{subfig:WkxSensVV}} \hfill \\
    \sidesubfloat[]{\hspace{-0.5cm}
        \begin{tikzpicture}
            \node (img)  {\includegraphics[width = 5.0cm]{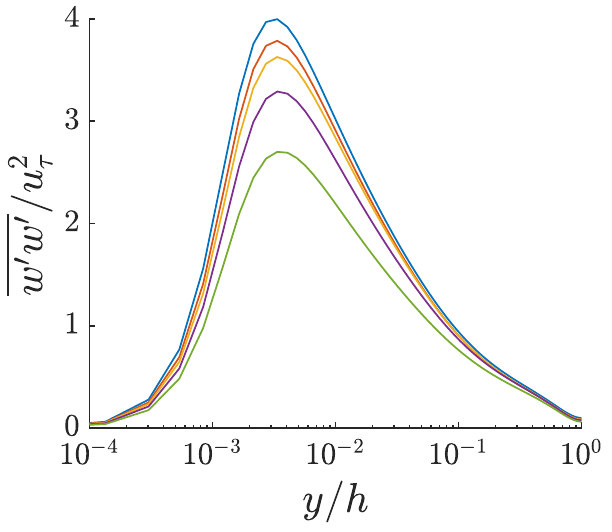}};
        \end{tikzpicture}
        \label{subfig:WkxSensWW}}
    \sidesubfloat[]{
        \begin{tikzpicture}
            \node (img)  {\includegraphics[width = 5.1cm]{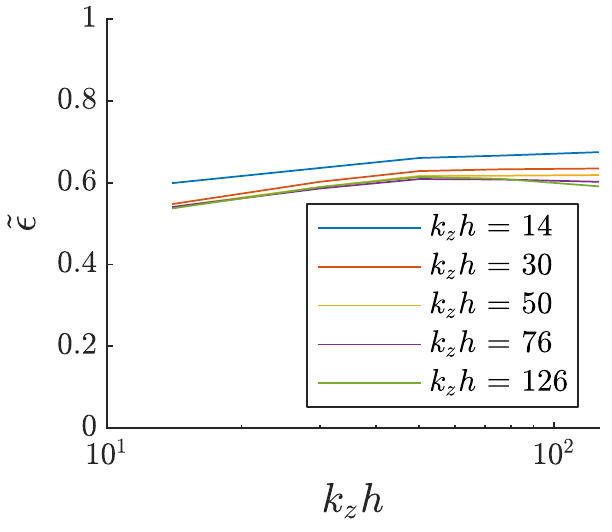}};
        \end{tikzpicture}
        \label{subfig:WkxSensTwoDspec}}
    \end{center}
    \caption{The sensitivity of the DQLA to the choice of streamwise weighting for (a) streamwise; (b) wall-normal; (c) spanwise turbulence intensity; and (d) total error in the two-dimensional, normalised spectra as defined in \eqref{eq:NormMin} with the streamwise weighting applied at different $k_zh$. Here the colour corresponds to the selected $k_zh$ result used for the self-similar streamwise weighting.}\label{fig:sensitivity}
\end{figure}

To check the sensitivity of the DQLA to the choice of self-similar streamwise weighting $W_{k_x}(k_x/k_z)$, the DQLA was performed at $Re_\tau \approx 5200$ with the different weights from figure \ref{fig:selfSimWeighting}. The turbulence intensity profiles are shown in figures. \ref{fig:sensitivity}(a-c). The streamwise turbulence intensity (figure \ref{subfig:WkxSensUU}) is relatively insensitive to the choice of the streamwise self-similar weight, while both the wall-normal and spanwise intensities (figures \ref{subfig:WkxSensVV},\ref{subfig:WkxSensWW}) tend to decrease with $k_zh$. 
The use of the $k_zh = 30$ weight is justified considering here that the wall-normal and spanwise turbulence intensity profiles are much less sensitive for the $k_zh = 14 - 50$ weights. Since these wavenumbers are mainly associated with the logarithmic layer, where self-similarity is expected to hold, the different weights from solving \eqref{eq:NormMin} lead to similar results in a DQLA. The sensitivity to the choice of self-similar streamwise weighting $W_{k_x}(k_x/k_z)$ is also examined in figure \ref{fig:sensitivity}(d), where the total errors between the normalised two-dimensional spectra for fixed spanwise length scales are examined using the weights for different $k_zh$ in figure \ref{fig:selfSimWeighting}, i.e.
\begin{equation}\label{eq:NormMinNormalised}
    \sum_s\norm{\frac{\bm \Phi^{\mathrm{DNS}}_{s}}{\norm{\bm \Phi^{\mathrm{DNS}}_{s}}_Q} 
    - \frac{\bm \Phi^{\mathrm{DQLA}}_{s}(W_{r,k_x})}{\norm{\bm \Phi^{\mathrm{DQLA}}_{s}}_Q}}_Q,
\end{equation}
for $s = \{uu,vv,ww,uv\}$. Figure \ref{fig:sensitivity}(d) shows that all weights produce a qualitatively similar trend. This justifies the use of the weights as self-similar weights at the other $k_zh$. The normalised spectra produce approximately the same total errors, giving the same approximate statistical structure of the resulting spectra. 

\bibliographystyle{jfm}
\bibliography{references}

\end{document}